\documentclass[12pt]{article}
\usepackage{graphicx}
\usepackage{epsfig}
\usepackage{amsfonts}
\usepackage[lofdepth,lotdepth]{subfig}
\usepackage{url}
\usepackage{setspace}
\usepackage{multirow}
\usepackage{array}
\usepackage{soul}
\usepackage{arydshln}

\begin{document}

\title{A bonding model of entanglement for $N$-qubit graph states}
\author{Mordecai Waegell \footnote{caiw@wpi.edu} \\ \it{ Physics Department, Worcester Polytechnic Institute} \\ {\it Worcester, MA, USA}}
\maketitle

\abstract{The class of entangled $N$-qubit states known as graph states, and the corresponding stabilizer groups of $N$-qubit Pauli observables, have found a wide range of applications in quantum information processing and the foundations of quantum mechanics.  A review of the properties of graph states is given and core spaces of graph states are introduced and discussed.  A bonding model of entanglement for generalized graph states is then presented, in which the presence or absence of a bond between two qubits unequivocally specifies whether or not they are entangled.  A physical interpretation of these bonds is given, along with a characterization of how they can be created or destroyed by entangling unitary operations and how they can be destroyed by local Pauli measurements.  It is shown that local unitary operations do not affect the bond structure of a graph state, and therefore that if two graph states have nonisomorphic bond structures, then local unitary operations and/or reordering of qubits cannot change one into the other.  Color multigraphs are introduced to depict the bond structures of graph states and to make some of their properties more apparent.}

\subsection{Introduction}

One of the most counterintuitive and fascinating features of quantum physics is the phenomenon of entanglement.  Entanglement is a special type of physical connection that can exist between different particles, such that certain local interactions with just one of the particles seem to have an instantaneous effect on all of the others \cite{EPR,Bell1,GHZ}, calling into question the very idea that we can even think of them as distinct physical objects.  In the mathematical language of quantum mechanics, entanglement arises because the Hilbert space of multiple particles admits states that cannot be factored into the direct product of independent states for each particle.  Furthermore we are able to realize these entangled states experimentally by beginning with a product state and performing particular unitary operations to `rotate' it into the entangled part of its Hilbert space.  This definition is straightforward, but it does not nail down the nature of the physical connection between the `distinct' particles in an entangled state.  In fact, simply knowing that a set of more than a few particles \cite{wootters1998entanglement, coffman2000distributed, wootters2001entanglement} are mutually entangled generally tells us little about the details of how that specific quantum state will behave.

It is the purpose of this letter to present a {\it bonding} model of entanglement within the $N$-qubit Pauli group, in which a given pair of qubits are bonded if and only if they are entangled, and which furthermore captures the essential details of how different types of multiqubit graph states behave under unitary evolution and local Pauli measurement.  While not particularly novel, we hope that this model represents a useful review and summary of much work that has already been done to develop graph states and their properties \cite{hein2006entanglement}.

Graph states and other geometric structures related to the $N$-qubit Pauli group have been the focus of much of the recent development in quantum information processing, in which entanglement, contextuality, and nonlocality all have interesting roles to play \cite{ekert1991quantum, bennett1996mixed, calderbank1997quantum, steane1996multiple, cleve1998quantum, ekert1998quantum, raussendorf2001one, raussendorf2003measurement, lanyon2013measurement, lanyon2013experimental, Walther2005, W_Primitive, WA_Nqubits}.  Within this group, entanglement alone lies at the heart of all of these nonclassical behaviors, and so it must hold the key to understanding all of the uniquely quantum phenomena that can be applied as resources for information processing.

It has also become apparent that generalizations of graph theory are a very natural framework in which to examine entanglement of many types \cite{YuOhGraphStates, Cabello_Pentagons, cabello2013exclusivity, coecke2010compositional, WaegellThesis, WA_3qubits, pavivcic2010new, elliott2008graphical, guhne2014entanglement} along with some of its nonclassical consequences, and in that vein we will introduce color multigraphs that generalize the usual graph generators from which graph states take their name.

The goal of the model we present here is to provide a useful means to visualize this class of entangled states and to make their general behavior more intuitive.

 The general features of the model are as follows.  We define entanglement-bonds between specific pairs of qubits within an entangled graph state.  These bonds are only created by the action of entangling multiqubit unitary operations, while single-qubit (local) unitary operations never alter them.  Multiqubit unitary operations can also naturally be used to destroy these bonds.  The bonds are also destroyed (and never created) by the action of local Pauli measurement operations.  Because local unitary operations never change the structure of bonds in a given state, each of the {\it lu}-inequivalent (local unitary) classes of $N$-qubit graph state is uniquely identified by its bond structure.  We will furthermore show that the structure of these bonds completely reveals the behavior of each state under unitary evolution and local Pauli measurements.

The remainder of this paper contains an introduction to several key components, and then brings them together to convey a complete picture of the bonding model.

\subsection{Pauli Measurements and Stabilizer Groups}

Before we begin with the discussion of bonds, let us first consider the effect of a local Pauli measurement on a graph state.

For each $N$-qubit graph state, there exists a stabilizer group of $2^N$ mutually commuting observables from the $N$-qubit Pauli group (including the $N$-qubit identity, which we will always omit), generated by $N$ independent observables from within the group.  A general graph state (stabilizer state) is defined by a fixed set of eigenvalues corresponding to each of the observables in the stabilizer (which are not all independent).  Throughout this paper we intentionally avoid the usual state vector representation and work entirely in terms of stabilizer groups and eigenvalues.

The effect of a projective Pauli measurement on a graph state is simply to truncate its original stabilizer group, by eliminating all elements that do not commute with the elements of the measurement stabilizer group.  For our purposes we will only be considering the effect of local Pauli measurements, since Pauli measurements on more qubits can always be decomposed into a multiqubit unitary (which can generally change one graph state to another) followed by local measurements.

If we perform a local Pauli measurement $A$ on the $i$th qubit in a maximally entangled graph state (i.e, $A$ is chosen from the set \{$Z_i$, $X_i$, $Y_i$\}), we simply discard all elements of its stabilizer group that do not commute with $A$ to obtain the outcome stabilizer $S_A$.  This is just a restatement of the fact that when one measures an observable, only eigenstates of that observable can be obtained as outcomes.  It is perhaps somewhat less obvious that the portion of the observables in $S_A$ that belong to other qubits are not affected by the measurement, and the outcome state simply remains a joint eigenstate of these observables.  To be specific, if we measure $Z_i$, we keep all observables in the stabilizer that explicitly contain $Z_i$ or $I_i$, and discard all observables that explicitly contain $X_i$ or $Y_i$.  Next we update the eigenvalues of the remaining observables based on the outcomes of the measurement - if the outcome is $+1$, all eigenvalues are unchanged, and if it is $-1$, then the eigenvalue gets flipped for each remaining observable that explicitly contains $A$ ($Z_i$ in the example).  For compactness, the measured qubit is then also removed from the stabilizer.  This qubit is still physically present, but no longer entangled with the others.  Finally the post-measurement graph state is defined by the updated eigenvalues of the now-truncated stabilizer group, as shown in Figs. \ref{GHZ3PX} and \ref{GHZ3PZ}.  We will use this interpretation of measurements as simply truncating the stabilizer group of the state to define how bonds are destroyed by the measurement process.  Additional examples of this truncation for various Pauli measurements on two specifically chosen graph states are clearly illustrated in Figs. \ref{ClusterPZ}, \ref{ClusterPX}, \ref{Cluster5PZ1}, \ref{Cluster5PX1}, \ref{Cluster5PZ5}, and \ref{Cluster5PY5}.

This truncation method can also be naturally generalized to any $N$-qubit stabilizer group that is closed under multiplication, even one with $M<N$ independent generators that defines an eigenbasis of rank-$r$ projectors ($r = 2^{N-M}$).  In this case, it reduces an $N$-qubit rank-$r$ projector into an $(N-1)$-qubit rank-$r$ projector.

Considering this truncation process, we can see that a given stabilizer group contains, and is essentially composed from, all of the different stabilizers to which it can reduce under local measurements.  For maximally entangled states, this amounts to a complete map of how the entanglement breaks down under successive local Pauli measurements.

\begin{figure}[h!]
\caption{The truncation process for a projective measurement $X_1$ on the 3-qubit GHZ state, leaving the remaining qubits in a Bell state.}
\centering
\subfloat[][$P_{\pm X_1} |\psi_{GHZ^3}\rangle$]{
\begin{tabular}{c|ccc}
$\lambda_1$ & $\textbf{X}$ & $X$ & $X$\\
$\lambda_2$ & $\textbf{X}$ & $Y$ & $Y$\\
\st{$\lambda_3$} & \st{$Y$} & \st{$X$} & \st{$Y$}\\
\st{$\lambda_4$} & \st{$Y$} & \st{$Y$} & \st{$X$}\\
$\lambda_5$ & $\textbf{I}$ & $Z$ & $Z$\\
\st{$\lambda_6$} & \st{$Z$} & \st{$I$} & \st{$Z$}\\
\st{$\lambda_7$} & \st{$Z$} & \st{$Z$} & \st{$I$}\\
\end{tabular}\label{GHZ3_PX}}
\qquad
\subfloat[][$|\psi_{Bell}\rangle$]{
\begin{tabular}{c|cc}
$\pm \lambda_1$  & $X$ & $X$\\
$\pm \lambda_2$  & $Y$ & $Y$\\
$\lambda_5$  & $Z$ & $Z$\\
\end{tabular}\label{GHZ3_APX}}
\label{GHZ3PX}
\end{figure}

\begin{figure}[h!]
\caption{The truncation process for a projective measurement $Z_1$ on the 3-qubit GHZ state, leaving the remaining qubits in a product state.}
\centering
\subfloat[][$P_{\pm Z_1} |\psi_{GHZ^3}\rangle$]{
\begin{tabular}{c|ccc}
\st{$\lambda_1$} & \st{$X$} & \st{$X$} & \st{$X$}\\
\st{$\lambda_2$} & \st{$X$} & \st{$Y$} & \st{$Y$}\\
\st{$\lambda_3$} & \st{$Y$} & \st{$X$} & \st{$Y$}\\
\st{$\lambda_4$} & \st{$Y$} & \st{$Y$} & \st{$X$}\\
$\lambda_5$ & $\textbf{I}$ & $Z$ & $Z$\\
$\lambda_6$ & $\textbf{Z}$ & $I$ &$Z$\\
$\lambda_7$ & $\textbf{Z}$ & $Z$ &$I$\\
\end{tabular}\label{GHZ3_PZ}}
\qquad
\subfloat[][$|Z\rangle \otimes |Z\rangle$]{
\begin{tabular}{c|c:c}
$\lambda_5$  & $Z$ & $Z$\\
$\pm \lambda_6$  & $I$ & $Z$\\
$\pm \lambda_7$  & $Z$ & $I$\\
\end{tabular}\label{GHZ3_APZ}}
\label{GHZ3PZ}
\end{figure}

\subsection{Core Spaces}

Before we move on to the bonding model, we will take a brief aside to discuss another interesting application of the compositional structure of stabilizer groups.  Consider that for a given stabilizer group and a corresponding set of eigenvalues, the observables that explicitly contain $I_i$ will never be discarded by a (general) local measurement on qubit $i$, nor will their eigenvalues ever be altered by such a measurement.  In this way, this set of observables form an invariant core, which can illustrate certain universal properties of the other qubits in the state when qubit $i$ is measured.  In particular, no measurement on qubit $i$ can ever result in a state that is less entangled than the core about qubit $i$.  To obtain the {\it core about qubit} $i$ in a graph state, one simply extracts the set of observables that contain $I_i$, along with their eigenvalues.

The core of a maximally entangled graph state is a rank-2 projector, or 2-dimensional {\it core space}, that contains all possible states to which the graph state can reduce when qubit $i$ is measured (in any basis, Pauli or not), as shown in Fig. \ref{ClusterCore}.  To see this consider that a maximally entangled $N$-qubit graph stabilizer has core stabilizers about each qubit that each contain $2^{N-2}$ elements (the proof of this is given in the Appendix).  The core stabilizer therefore contains only $N-2$ independent generators for its $N-1$ qubits, and thus defines an eigenbasis of rank-2 projectors.  Furthermore, all {\it lu}-inequivalent graph states for up to $N=12$ are known \cite{hein2006entanglement}, and for $N\geq5$ there exist graph states whose core spaces can be maximally entangled subspaces of all $N-1$ qubits, such as the 5-qubit pentagon state shown in Fig. \ref{PentagonCore}.

Special sets of Pauli observables from within graph stabilizer groups called {\it identity products} (IDs) were introduced in \cite{W_Primitive}, and are described using the compact symbol ID$M^N$ to denote a set of $M$ observables from within an $N$-qubit stabilizer group whose overall product is $\pm I$ (the $N$-qubit identity).  IDs are furthermore called {\it critical} if it is impossible to remove qubits and/or observables from the set in order to obtain a smaller ID.  Critical IDs always contain maximally entangled subsets of exactly $M-1$ independent generators, which are analogous to the usual graph generators for cases where $M=N+1$.  For cases where $M<N+1$, these subsets belong to several different maximally entangled stabilizer groups, and thus define maximally entangled subspaces of dimension $d=2^{N-M+1}$.

The 4-qubit cluster state (Figs. \ref{q4ClusterBonds}, \ref{ClusterPZ}, and \ref{ClusterPX}) is not symmetric under all exchanges of qubits, and there are in fact three distinct 4-qubit cluster states that are isomorphic, but locally inequivalent to one another.  Nevertheless, all three are eigenstates of the same critical ID$4^4$, and thus they all belong to the same maximally entangled 2-dimensional subspace, which we can call the 4-qubit {\it cluster space}.  For the 5-qubit pentagon state of Figs. \ref{q5Pentcounts} and \ref{PentagonCore}, all 5 core spaces are the cluster space.

Naturally we can also generalize this concept by looking at the core spaces of core spaces, which is identical to considering the core space of the original state about $n$ of its qubits instead of just one.  To obtain a generalized core stabilizer about $n$ qubits, we simply extract all observables from the original stabilizer that contain $I_i$ for all $n$ of those qubits.  The core spaces obtained in this way are then of maximum dimension $d=2^n$ (maximum because we must now consider cases where not every qubit in a core is entangled - see the Appendix).  These core spaces contain all states to which the original $N$-qubit graph state can reduce under arbitrary {\it local} measurements on those $n$ qubits.

If all critical IDs generate maximally entangled core spaces in the same way as the critical ID$4^4$ (which seems very likely), then our most extreme example of a core space is provided by the critical ID$7^{16}$ (Fig. 12c of \cite{W_Primitive}), which contains 6 independent generators for 16 qubits.  This would then be the maximally entangled 1024-dimensional core space of a maximally entangled graph state of at least 26 qubits, about at least 10 of its qubits.

Core spaces elucidate the entanglement persistency of a given graph state, including its Pauli persistency \cite{briegel2001persistent,BriegelGraphStates}.  The codespaces used in quantum error correction protocols \cite{NielsenChuang, barz2013demonstrating, waldherr2014quantum, bell2014experimental} are also naturally a special subset of all core spaces.  Core spaces are also of some foundational interest with regard to the underlying structure of entanglement, which is made more apparent by the bonding model.

\subsection{Maximally Entangled Subparts}

The most fundamental detail we need in order to define our bonding model is a broad method for answering the yes-or-no question of whether or not two specific qubits are entangled.  To do this, we exploit the commutation properties of the entangled set of observables in the stabilizer group of a maximally entangled graph state.  Such stabilizers have the important property that their observables cannot all be simultaneously factored into the tensor products of observables for two subparts such that all of the factors of each individual subpart mutually commute \cite{W_Primitive}.  This means that we can determine which qubits are entangled within a given stabilizer simply by identifying all minimal sets of qubits for which that entire sub-stabilizer mutually commutes.  In any graph stabilizer group, these sets are always mutually disjoint, and thus give a decomposition of the stabilizer into distinct entangled subparts, as shown by the vertical dashed lines in the post-measurement stabilizer groups of Figs. \ref{GHZ3_APZ}, \ref{Cluster_APX}, \ref{Cluster5_APZ1}, and \ref{Cluster5_APY5}.

\subsection{Entanglement Bonds}

Now that we can answer the broad question of whether or not a given pair of qubits are entangled, we have all of the necessary ingredients to construct our bond model.  A bond is a specific relation between 2 qubits and 2 observables within a stabilizer group, with a simple physical interpretation: {\it If two qubits are bonded, then it is always possible to obtain a Bell state of those two qubits by performing suitable local Pauli measurements on other qubits}.  The proof of this interpretation is given below.

For each pair of qubits and each pair of observables in the set, a bond exists if and only if both of the following conditions are met:  First, the two qubits must belong to the same entangled subpart of the stabilizer.  Second, for qubits $i$ and $j$ and $N$-qubit Pauli observables $A$ and $B$, $A_i$ must anticommute with $B_i$ and $A_j$ must anticommute with $B_j$.  It is important to consider the complete stabilizer group rather than just a set of generators if one wishes to see the complete structure of bonds.

\begin{figure}[h!]
\centering
\caption{(Color Online)  The color multigraph for the 2-qubit Bell state.  Each observable is assigned a color, and each edge is a mix of two such colors.  A local Pauli measurement causes deletion of 2 colors, and thus all 3 edges are also deleted, leaving an unentangled single-qubit state.}
\includegraphics[width=3in]{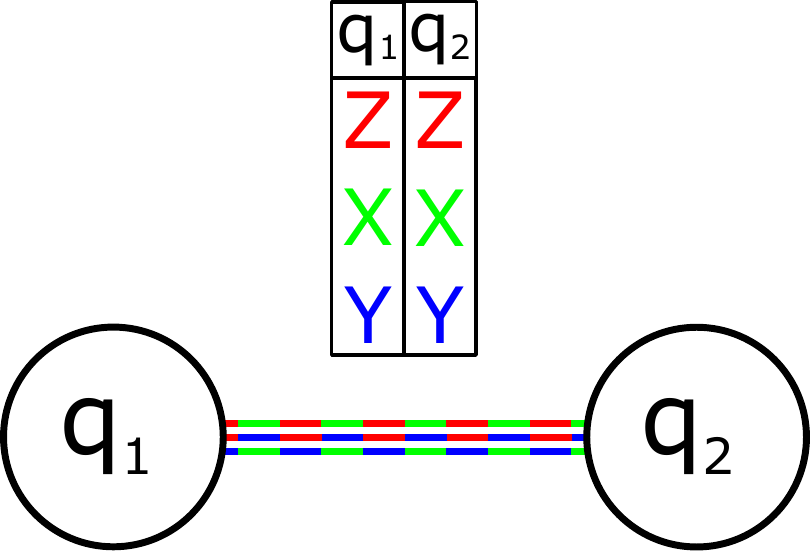}
\label{q2Bonds}
\end{figure}

\begin{figure}[h!]
\centering
\caption{(Color Online)  The color multigraph for the 3-qubit GHZ state.  Each observable is assigned a color, and each edge is a mix of two such colors.  A local Pauli measurement causes deletion of certain colors, and all edges that include those colors.}
\includegraphics[width=3.5in]{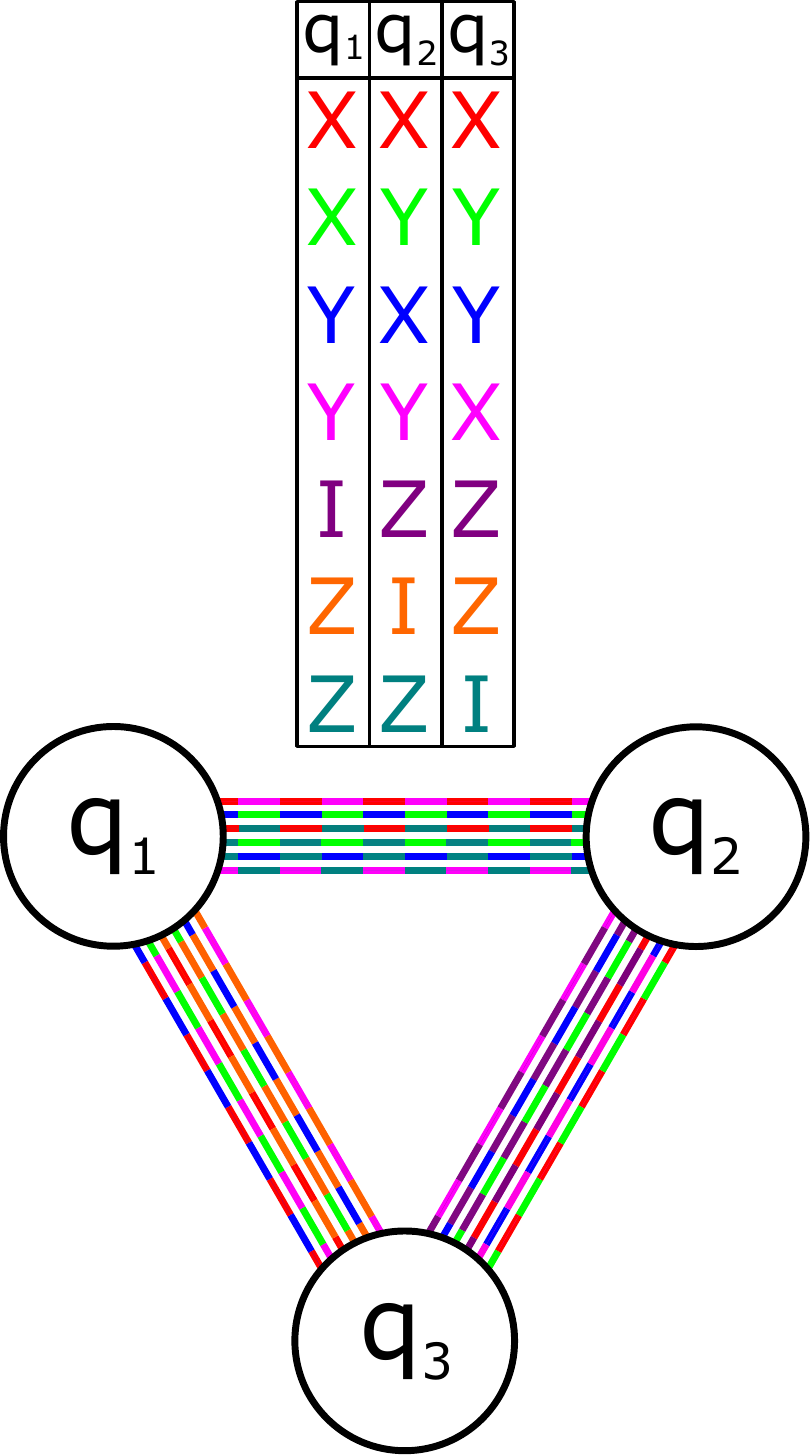}
\label{q3Bonds}
\end{figure}

Clearly there can be many bonds between a given pair of qubits, and so the bond structure diagram, in which each qubit is a vertex and each bond an edge, is an $N$-vertex color multigraph.  We assign a color to each observable of the stabilizer group, and then a bond is represented by an edge that is a mix of two of these colors ($A$ and $B$), connecting two of those qubits ($i$ and $j$).  Unfortunately, the number of bonds and colors in these multigraphs grows too rapidly with $N$ to make it practical to draw them for $N>3$ (see Figs. \ref{q2Bonds} and \ref{q3Bonds}), and so visualizing the larger cases is left to the reader's imagination.
\begin{figure}[h!]
\caption{Compacted multigraphs of all nonisomorphic bond structures for up to $N=5$, using the graph state numbering given in \cite{hein2006entanglement}.  The qubits in \protect\subref{q4ClusterBonds} and \protect\subref{q5ClusterBonds} have been ordered to match the stabilizers given in other figures.}
\centering
\subfloat[][No. 3: $|\psi_{GHZ^4}\rangle$]{
\includegraphics[width=1.5in]{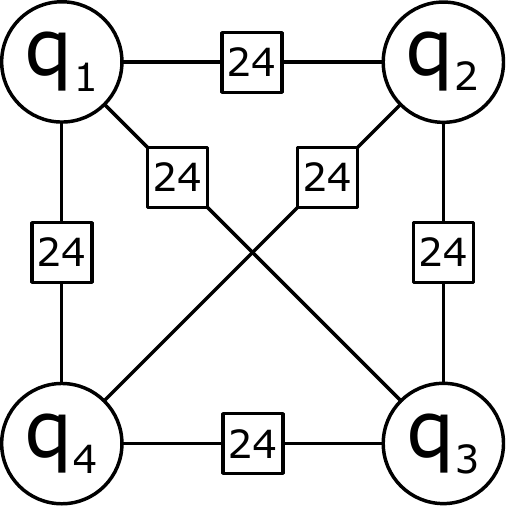}
}
\qquad
\subfloat[][No. 4: $|\psi_{Cluster^4}\rangle$]{
\includegraphics[width=1.5in]{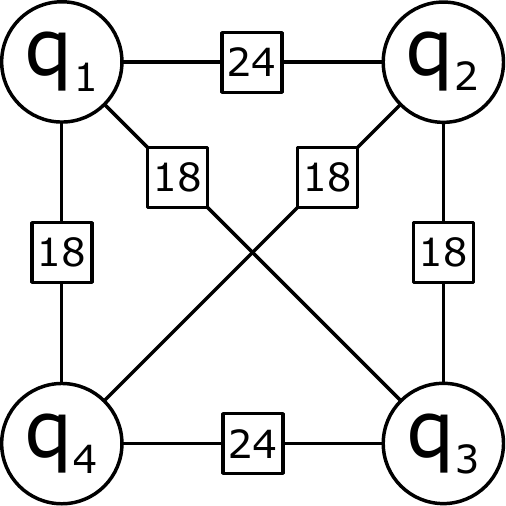}
\label{q4ClusterBonds}}
\qquad
\subfloat[][No. 5: $|\psi_{GHZ^5}\rangle$]{
\includegraphics[width=1.5in]{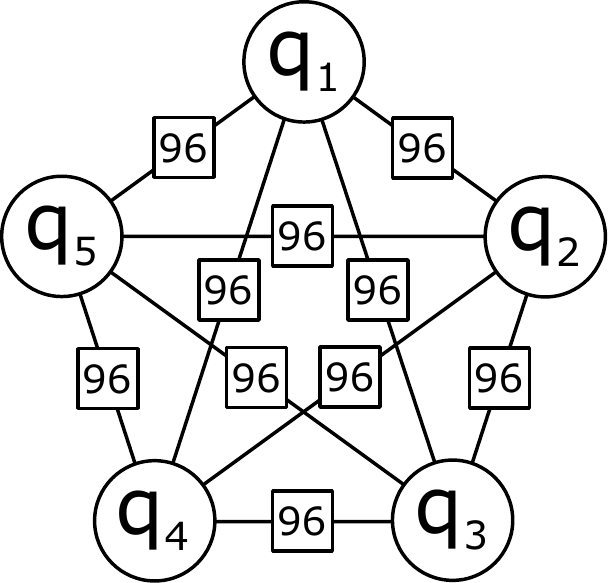}
}
\qquad
\subfloat[][No. 6: $|\psi_{Cluster_B^5}\rangle$]{
\includegraphics[width=1.5in]{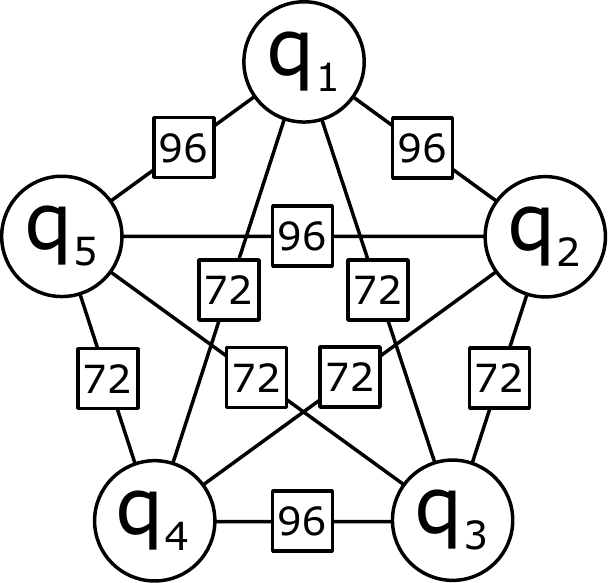}
}
\qquad
\subfloat[][No. 7: $|\psi_{Cluster^5}\rangle$]{
\includegraphics[width=1.5in]{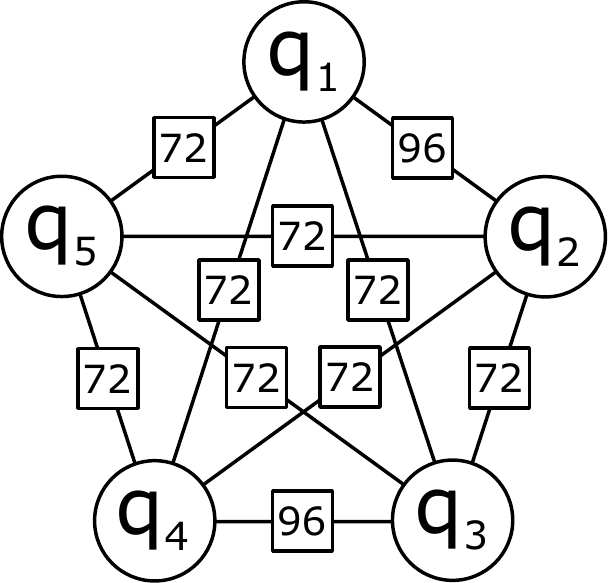}
\label{q5ClusterBonds}}
\qquad
\subfloat[][No. 8: $|\psi_{Pentagon^5}\rangle$]{
\includegraphics[width=1.5in]{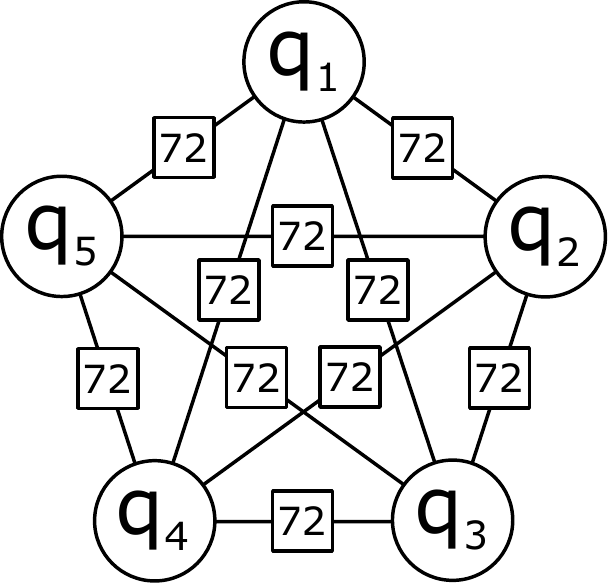}
\label{q5Pentcounts}}\label{q45bondcounts}
\end{figure}

It is worth noting that the standard graph state generators are also constructed using bonds of the form $\{A_i=X, B_i = Z, A_j = Z, B_j = X\}$, but in such a way that within the set of generators there is no more than one bond (edge) between each pair of vertices (thus the bond diagram is a graph rather than a multigraph).  Generating the full stabilizer group then generates the complete bond structure multigraph.  There are various ways to extract a set of independent generators with at most one bond between each pair of qubits from within a given stabilizer group, which is why we find that numerous nonisomorphic graphs generate the same stabilizer group (these are the so-called local-complementation orbits produced by local Clifford operations) \cite{hein2006entanglement, BriegelGraphStates}.

A key property of these bonds is that they are invariant under the effect of local unitary operations, and thus the only unitary operations that can create or destroy them are entangling multiqubit operations.  To see this, consider that the effect of a local unitary is simply to map the Pauli basis ($Z,X,Y$) into a transformed basis ($Z',X',Y'$), and thus the anticommutation requirement for bonds is never affected.  This also means that if two different entangled states are inequivalent under local unitary operations and/or exchanges of qubits, then their respective bond structures, and thus their respective color multigraphs, must be nonisomorphic (where morphisms now include exchanges of colors).  Thus the question of identifying all distinct types of entanglement among graph states can be recast as a generalization of the graph isomorphism problem (which has been solved for up to $N=12$, see Table IV of \cite{hein2006entanglement}).

Figure \ref{q45bondcounts} shows compacted forms of the color multigraphs for all of nonisomorphic bond structures for up to $N=5$, in which each edge now indicates only the count of bonds between each pair of qubits, but no longer contains any detail about how the bonds break down under measurement.  Up to $N=5$ these counts are actually sufficient to distinguish one multigraph from another, but this not generally true.  For larger $N$ one can find two or more nonisomorphic multigraphs (i.e. corresponding to states with different types of entanglement) that nevertheless have the same bond counts.

The most elementary operation that creates bonds is the 2-qubit controlled-$Z$ (CZ) gate, whose action on the observables of a stabilizer group is quite straightforward \cite{hein2006entanglement} (Fig. \ref{CZ}, and see also Chapter 10.5.2 of \cite{NielsenChuang} for a general discussion of how unitaries act on stabilizer groups), and indeed all $N$-qubit unitary operations can be decomposed into a mix of CZ gates and local unitary operations (i.e. CZ is locally equivalent to CNOT).  For graph states defined in the conventional way, the CZ operations create or destroy a single bond (edge) between a given pair of qubits (vertices), and indeed all graph states can be constructed in this way, by starting with a product state and applying a single CZ gate for each edge in the graph.

However, the action of CZ on the complete bond structure is somewhat more subtle.  If a CZ gate acts on two qubits that already belong to the same entangled group, then it is possible for this unitary to create and/or destroy bonds throughout that group - depending on the details.  If however, a CZ gate acts on two qubits that belong to two different entangled groups in such a way that it creates a bond between those two qubits, it then also creates bonds between every qubit in one group to every qubit in the other group (see Figs. \ref{CZP} and \ref{CZB}).

Finally let us consider the effect of local Pauli measurements on the bond structure of a given graph state.  As we discussed, local Pauli measurements cause the stabilizer group to be truncated, and this in turn causes some bonds to be destroyed, such that the remaining bond structure is the correct bond structure of the truncated stabilizer.  This requires two different mechanisms for removing bonds.  First, any bond between one or more of the discarded observables is destroyed - and this always removes all of the bonds connected to the measured qubit.  Second, the truncated stabilizer may decompose into more entangled subparts than before, and any bond between qubits that no longer belong to the same entangled subpart is also destroyed.  Note that only deletion of bonds is required to characterize the measurement process within this bond model, and so a Pauli measurement simply excises a particular piece from within the original bond structure of the state, just as it excises a particular piece from within the state's vector representation.\newline

We now explicitly show how the CZ gate acts on the observables of a stabilizer, and demonstrate how it creates bonds between different maximally entangled subparts of a stabilizer.
\begin{figure}[h!]
\caption{Action of controlled-Z (CZ), $C_Z^{a,b}$ operation on stabilizer elements, with $a$ the control qubit and $b$ the target qubit.}
\centering
\begin{tabular}{c}
$C_Z^{1,2} (XI) C_Z^{1,2} = XZ$ \\
$C_Z^{1,2} (IX) C_Z^{1,2} = ZX$ \\
$C_Z^{1,2} (ZI) C_Z^{1,2} = ZI$ \\
$C_Z^{1,2} (IZ) C_Z^{1,2} = IZ$ \\
\end{tabular}\label{CZ}
\end{figure}
It is important to note that unlike CNOT, CZ is symmetric under exchange of the control and target qubits.  It is also important to note that all $C_Z^{a,b}$ mutually commute for all values of $a$ and $b$, and thus the order in which they are applied to build up the edges of a conventional graph state is arbitrary.

Consider first the case that a CZ gate acts on a qubit from within a subpart that is already maximally entangled, and another qubit that is previously not entangled at all.  Because the stabilizer of any such product state is the tensor-set product of the stabilizers of the two different (unentangled) systems, it is guaranteed to contain elements of $A$ in Fig. \ref{CZAP} (up to local unitaries), neglecting additional observables and/or qubits that may belong to the complete stabilizer.
\begin{figure}[h!]
\caption{$A$ is an elementary piece of any stabilizer in which qubits 1 and 2 are bonded, while qubit 3 is not entangled at all (as denoted by the vertical dashed line).  The CZ operation on qubits 2 and 3 of $A$ to obtain $B$, in which all 3 qubits are now entangled.}
\centering
\subfloat[][$A$]{
\begin{tabular}{cc:c}
$X$ & $Z$ & $I$\\
$Z$ & $X$ & $I$\\
$X$ & $Z$ & $X$\\
$Z$ & $X$ & $X$\\
\end{tabular}\label{CZAP}}
\qquad
\subfloat[][$B = C_Z^{2,3} A C_Z^{2,3}$]{
\begin{tabular}{ccc}
$X$ & $Z$ & $I$\\
$Z$ & $X$ & $Z$\\
$X$ & $I$ & $X$\\
$Z$ & $Y$ & $Y$\\
\end{tabular}\label{CZBP}}
\label{CZP}
\end{figure}
There are two important properties that be seen by considering the effect of entangling the existing bond in Fig. \ref{CZAP} with a new qubit.  First, it is easy to see that every pair of qubits in Fig. \ref{CZBP} are now bonded (even though we did not act on qubit 1).  Second, it easy to see that a projective measurement $Z_3$ will truncate off the last two observables, leaving behind the original bond.

Next, consider the case that CZ acts on two qubits that each belong a different maximally entangled subpart of the stabilizer.  Now the stabilizer is guaranteed to contain the elements of $A$ in Fig. \ref{CZAB} (up to local unitaries), neglecting additional observables and/or qubits that may belong to the complete stabilizer.
\begin{figure}[h!]
\caption{$A$ is an elementary piece of any stabilizer in which qubits 1 and 2 are bonded, and qubits 3 and 4 are bonded, but there is no entangledment between the two groups (as denoted by the vertical dashed line).  The CZ operation on qubits 2 and 3 of $A$ to obtain $B$, in which all 4 qubits are now entangled.}
\centering
\subfloat[][$A$]{
\begin{tabular}{cc:cc}
$X$ & $Z$ & $Z$ & $X$\\
$Z$ & $X$ & $Z$ & $X$\\
$X$ & $Z$ & $X$ & $Z$\\
$Z$ & $X$ & $X$ & $Z$\\
\end{tabular}\label{CZAB}}
\qquad
\subfloat[][$B = C_Z^{2,3} A C_Z^{2,3}$]{
\begin{tabular}{cccc}
$X$ & $Z$ & $Z$ & $X$\\
$Z$ & $X$ & $I$ & $X$\\
$X$ & $I$ & $X$ & $Z$\\
$Z$ & $Y$ & $Y$ & $Z$\\
\end{tabular}\label{CZBB}}
\label{CZB}
\end{figure}
Again, we now find bonds between every pair of qubits in Fig. \ref{CZBB}.  And again, local measurements can remove the two new qubits and restore the (either) original bond (for example, by measuring $Z_3$ and $X_4$).

The two cases above give a completely general (inductive) argument for using a CZ gate to entangle two parts of a stabilizer group that were not previously entangled (regardless of size), showing that every pair of qubits will become bonded within any maximally entangled subpart.  Another way to say this is that because every maximally entangled graph state corresponds to a connected graph, it also corresponds to a fully connected multigraph - though the number of bonds (edges) may not be symmetric for all pairs of qubits (vertices).

These two cases also show that there is always a set of local measurements that effectively reverse the new entanglement, allowing one to extract the original bond.  This proves our statement about the physical interpretation of bonds, because one can use such a set of local operations on other qubits to extract a bond between {\it any} two qubits from within a maximally entangled stabilizer, leaving those qubits in a Bell state.

We must stress that the projective measurement feature of our model only applies to Pauli measurements ($Z,X,Y$).   Making a measurement in another basis results in a non-graph state that falls outside the scope of this model.  There is however no requirement that the bases for each qubit be aligned in any way, and thus ($Z_1,X_1,Y_1$) and ($Z_2,X_2,Y_2$) can be completely unrelated bases in defining the general graph state (this is the complete freedom of local unitary operations), but then these same bases are the only ones that can be measured as `Pauli' measurements within the bonding model.

The exception to the above rule is that the bond structure of the core space about qubit $i$ will never be affected by any local measurement on that qubit, regardless of what basis it is measured in.  In general these measurements still create a state that falls outside the bond model, but that state nevertheless contains the bond structure of the core, and falls within the core space.  This of course generalizes to the core space of a state about more than one of its qubits - we can always identify the core bonds (if any) that cannot be affected by local measurements on those particular qubits.

\begin{figure}[h!]
\centering
\caption{(Color Online)  The color multigraph for the 4-qubit cluster space.  The observables are shown in a different order than in Fig. \ref{PentagonCore} in order to make the symmetry of the structure more apparent.}
\includegraphics[width=3.5in]{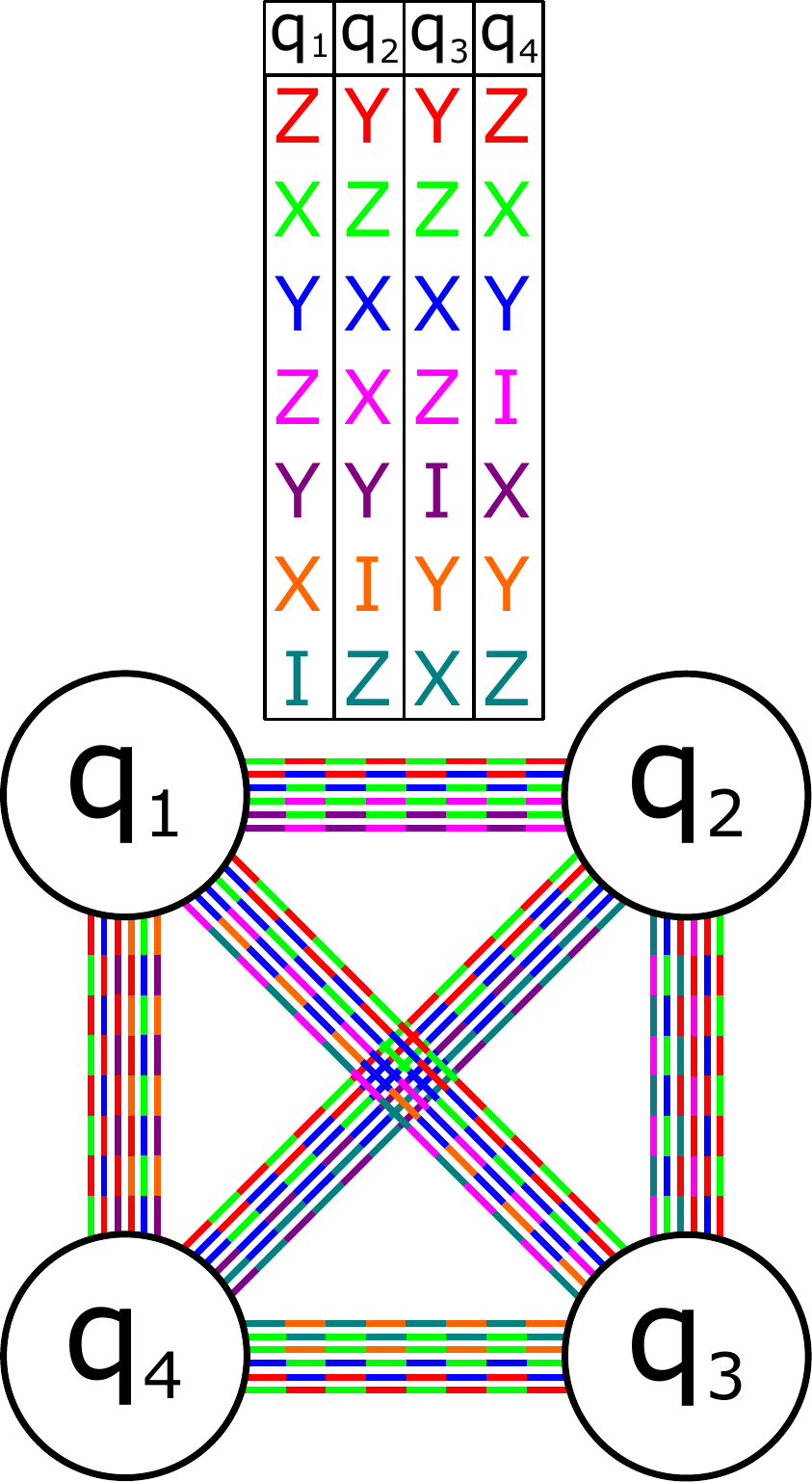}
\label{ClusterSpaceBonds}
\end{figure}

The bond structure of the cluster space is shown in Fig. \ref{ClusterSpaceBonds}.  The cluster space is the core space of the 5-qubit pentagon state about any one of its qubits, and therefore any local measurement (Pauli or not) on it results in a state within this space, and with these bonds.  This also demonstrates that the underlying entanglement structure is much more apparent in the complete bond picture than it is in the usual graph picture, which can only accommodate rank-1 projectors.

\subsection{Conclusions}

Let us review the properties of the bonding model we have defined:
\begin{itemize}
  \item We have explicitly defined entanglement bonds within the stabilizer groups of generalized graph states, and given their straightforward physical interpretation.
  \item We have shown how each graph state can be represented by a single color multigraph, and that these color multigraphs are nonisomorphic if and only if the graph states are inequivalent under local unitary operations and/or exchanges of qubits.
  \item We have discussed in terms of local unitary operations and 2-qubit controlled-Z gates, how general multiqubit unitary operations can alter the bond structure, changing one graph state into another.
  \item We have specified how local Pauli measurements truncate the bond structure, reducing one graph state into another.
  \item We have identified the core spaces of graph states about one or more of their qubits, and thus the core bond structure that is never affected by local measurements on those qubits.
\end{itemize}

It is our hope that this review of graph states and stabilizer groups, along with the accompanying bond model, will help to provide a broader and more general view of the structure of graph states and their underlying behavior.  In particular we would like to stress the remarkable structure and foundational significance of maximally entangled spaces, which have already found practical applications in quantum error correction \cite{barz2013demonstrating, waldherr2014quantum, bell2014experimental}, and which promise to have more interesting lessons in store.

As a final note, it would be interesting to determine if entangled states such as the W state, that do not belong to the $N$-qubit Pauli group \cite{verstraete2002four, miyake2005distillation}, are amenable to such a model, and if so, how well our intuitions will translate from one case to another.

\textbf{Acknowledgement:}  My thanks to P.K. Aravind for many useful discussions as I fleshed out the details of this work, and also to my anonymous referee who made several useful suggestions.

\bibliographystyle{ieeetr}
\bibliography{Entanglement_Bonds.bbl}

\begin{thebibliography}{10}

\bibitem{EPR}
A.~Einstein, B.~Podolsky, and N.~Rosen, ``Can quantum-mechanical description of
  physical reality be considered complete?,'' {\em Physical Review}, vol.~47,
  no.~10, p.~777, 1935.

\bibitem{Bell1}
J.~Bell, ``On the {E}instein-{P}odolsky-{R}osen paradox,'' {\em Physics},
  vol.~1, no.~3, pp.~195--200, 1964.

\bibitem{GHZ}
D.~Greenberger, M.~Horne, and A.~Zeilinger, ``Going beyond {B}ell's theorem,''
  {\em Bell's theorem, quantum theory, and conceptions of the universe},
  vol.~37, pp.~69--72, 1989.

\bibitem{wootters1998entanglement}
W.~K. Wootters, ``Entanglement of formation of an arbitrary state of two
  qubits,'' {\em Physical Review Letters}, vol.~80, no.~10, p.~2245, 1998.

\bibitem{coffman2000distributed}
V.~Coffman, J.~Kundu, and W.~K. Wootters, ``Distributed entanglement,'' {\em
  Physical Review A}, vol.~61, no.~5, p.~052306, 2000.

\bibitem{wootters2001entanglement}
W.~K. Wootters, ``Entanglement of formation and concurrence,'' {\em Quantum
  Information \& Computation}, vol.~1, no.~1, pp.~27--44, 2001.

\bibitem{hein2006entanglement}
M.~Hein, W.~D{\"u}r, J.~Eisert, R.~Raussendorf, M.~Nest, and H.-J. Briegel,
  ``Entanglement in graph states and its applications,'' {\em arXiv preprint
  quant-ph/0602096}, 2006.

\bibitem{ekert1991quantum}
A.~K. Ekert, ``Quantum cryptography based on {B}ell's theorem,'' {\em Physical
  Review Letters}, vol.~67, no.~6, pp.~661--663, 1991.

\bibitem{bennett1996mixed}
C.~H. Bennett, D.~P. DiVincenzo, J.~A. Smolin, and W.~K. Wootters,
  ``Mixed-state entanglement and quantum error correction,'' {\em Physical
  Review A}, vol.~54, no.~5, p.~3824, 1996.

\bibitem{calderbank1997quantum}
A.~R. Calderbank, E.~M. Rains, P.~W. Shor, and N.~J. Sloane, ``Quantum error
  correction and orthogonal geometry,'' {\em Physical Review Letters}, vol.~78,
  no.~3, pp.~405--408, 1997.

\bibitem{steane1996multiple}
A.~Steane, ``Multiple-particle interference and quantum error correction,''
  {\em Proceedings of the Royal Society of London. Series A: Mathematical,
  Physical and Engineering Sciences}, vol.~452, no.~1954, pp.~2551--2577, 1996.

\bibitem{cleve1998quantum}
R.~Cleve, A.~Ekert, C.~Macchiavello, and M.~Mosca, ``Quantum algorithms
  revisited,'' {\em Proceedings of the Royal Society of London. Series A:
  Mathematical, Physical and Engineering Sciences}, vol.~454, no.~1969,
  pp.~339--354, 1998.

\bibitem{ekert1998quantum}
A.~Ekert and R.~Jozsa, ``Quantum algorithms: entanglement--enhanced information
  processing,'' {\em Philosophical Transactions A}, vol.~356, no.~1743,
  p.~1769, 1998.

\bibitem{raussendorf2001one}
R.~Raussendorf and H.~J. Briegel, ``A one-way quantum computer,'' {\em Physical
  Review Letters}, vol.~86, no.~22, pp.~5188--5191, 2001.

\bibitem{raussendorf2003measurement}
R.~Raussendorf, D.~E. Browne, and H.~J. Briegel, ``Measurement-based quantum
  computation on cluster states,'' {\em Physical Review A}, vol.~68, no.~2,
  p.~022312, 2003.

\bibitem{lanyon2013measurement}
B.~Lanyon, P.~Jurcevic, M.~Zwerger, C.~Hempel, E.~Martinez, W.~D{\"u}r,
  H.~Briegel, R.~Blatt, and C.~Roos, ``Measurement-based quantum computation
  with trapped ions,'' {\em Physical review letters}, vol.~111, no.~21,
  p.~210501, 2013.

\bibitem{lanyon2013experimental}
B.~Lanyon, M.~Zwerger, P.~Jurcevic, C.~Hempel, W.~D{\"u}r, H.~Briegel,
  R.~Blatt, and C.~Roos, ``Experimental violation of multipartite bell
  inequalities with trapped ions,'' {\em arXiv preprint arXiv:1312.4810}, 2013.

\bibitem{Walther2005}
P.~Walther, M.~Aspelmeyer, K.~J. Resch, and A.~Zeilinger, ``Experimental
  violation of a cluster state {B}ell inequality,'' {\em Physical Review
  Letters}, vol.~95, no.~2, p.~20403, 2005.

\bibitem{W_Primitive}
M.~Waegell, ``Primitive {N}onclassical {S}tructures of the {$N$}-qubit {P}auli
  {G}roup,'' {\em arXiv preprint arXiv:1310.3419 [quant-ph], (accepted by PRA
  Dec. 2013)}, 2013.

\bibitem{WA_Nqubits}
M.~Waegell and P.~K. Aravind, ``Proofs of the {K}ochen-{S}pecker theorem based
  on the {$N$}-qubit {P}auli group,'' {\em Phys. Rev. A}, vol.~88, p.~012102,
  Jul 2013.

\bibitem{YuOhGraphStates}
W.~Tang, S.~Yu, and C.~Oh, ``{G}reenberger-{H}orne-{Z}eilinger paradoxes from
  qudit graph states,'' {\em Physical Review Letters}, vol.~110, no.~10,
  p.~100403, 2013.

\bibitem{Cabello_Pentagons}
P.~Badzi\c{a}g, I.~Bengtsson, A.~Cabello, H.~Granstr{\"o}m, and J.-{\AA}.
  Larsson, ``Pentagrams and paradoxes,'' {\em Foundations of Physics}, vol.~41,
  no.~3, pp.~414--423, 2011.

\bibitem{cabello2013exclusivity}
A.~Cabello, M.~G. Parker, G.~Scarpa, and S.~Severini, ``Exclusivity structures
  and graph representatives of local complementation orbits,'' {\em Journal of
  Mathematical Physics}, vol.~54, p.~072202, 2013.

\bibitem{coecke2010compositional}
B.~Coecke and A.~Kissinger, ``The compositional structure of multipartite
  quantum entanglement,'' in {\em Automata, Languages and Programming},
  pp.~297--308, Springer, 2010.

\bibitem{WaegellThesis}
M.~Waegell, {\em Nonclassical {S}tructures within the $N$-qubit {P}auli
  {G}roup}.
\newblock PhD thesis, Worcester Polytechnic Institute, 2013.
\newblock http://arxiv.org/abs/1307.6264.

\bibitem{WA_3qubits}
M.~Waegell and P.~K. Aravind, ``Proofs of the {K}ochen--{S}pecker theorem based
  on a system of three qubits,'' {\em J. Phys. A: Math. and Theor.}, vol.~45,
  p.~405301, 2012.

\bibitem{pavivcic2010new}
M.~Pavi{\v{c}}i{\'c}, N.~D. Megill, and J.-P. Merlet, ``New {K}ochen-{S}pecker
  sets in four dimensions,'' {\em Physics Letters A}, vol.~374, no.~21,
  pp.~2122--2128, 2010.

\bibitem{elliott2008graphical}
M.~B. Elliott, B.~Eastin, and C.~M. Caves, ``Graphical description of the
  action of clifford operators on stabilizer states,'' {\em Physical Review A},
  vol.~77, no.~4, p.~042307, 2008.

\bibitem{guhne2014entanglement}
O.~G{\"u}hne, M.~Cuquet, F.~E. Steinhoff, T.~Moroder, M.~Rossi, D.~Bru{\ss},
  B.~Kraus, and C.~Macchiavello, ``Entanglement and nonclassical properties of
  hypergraph states,'' {\em arXiv preprint arXiv:1404.6492}, 2014.

\bibitem{briegel2001persistent}
H.~J. Briegel and R.~Raussendorf, ``Persistent entanglement in arrays of
  interacting particles,'' {\em Physical Review Letters}, vol.~86, no.~5,
  p.~910, 2001.

\bibitem{BriegelGraphStates}
M.~Hein, J.~Eisert, and H.~Briegel, ``Multiparty entanglement in graph
  states,'' {\em Physical Review A}, vol.~69, no.~6, p.~062311, 2004.

\bibitem{NielsenChuang}
M.~A. Nielsen and I.~L. Chuang, {\em Quantum computation and quantum
  information}.
\newblock Cambridge University Press, 2010.

\bibitem{barz2013demonstrating}
S.~Barz, R.~Vasconcelos, C.~Greganti, M.~Zwerger, W.~D{\"u}r, H.~J. Briegel,
  and P.~Walther, ``Demonstrating elements of measurement-based quantum error
  correction,'' {\em arXiv preprint arXiv:1308.5209}, 2013.

\bibitem{waldherr2014quantum}
G.~Waldherr, Y.~Wang, S.~Zaiser, M.~Jamali, T.~Schulte-Herbr{\"u}ggen, H.~Abe,
  T.~Ohshima, J.~Isoya, J.~Du, P.~Neumann, {\em et~al.}, ``Quantum error
  correction in a solid-state hybrid spin register,'' {\em Nature}, 2014.

\bibitem{bell2014experimental}
B.~Bell, D.~Herrera-Mart{\'\i}, M.~Tame, D.~Markham, W.~Wadsworth, and
  J.~Rarity, ``Experimental demonstration of a graph state quantum
  error-correction code,'' {\em Nature communications}, vol.~5, 2014.

\bibitem{verstraete2002four}
F.~Verstraete, J.~Dehaene, B.~De~Moor, and H.~Verschelde, ``Four qubits can be
  entangled in nine different ways,'' {\em Physical Review A}, vol.~65, no.~5,
  p.~052112, 2002.

\bibitem{miyake2005distillation}
A.~Miyake and H.~J. Briegel, ``Distillation of multipartite entanglement by
  complementary stabilizer measurements,'' {\em Physical review letters},
  vol.~95, no.~22, p.~220501, 2005.

\end{thebibliography}

\subsection{Appendix: Structure of Maximally Entangled Stabilizer Groups}

Throughout this paper we have made implicit use of a basic fact about the structure of all maximally entangled stabilizer groups, and this is that every qubit in a maximally entangled stabilizer group always has an equal mix of $\{Z,X,Y,I\}$ among the observables to which it belongs.  This ensures that exactly half of the stabilizer group remains after truncation by a local Pauli measurement (as it must in order for the new stabilizer to be complete).  It also shows that the dimension of the core spaces of maximally entangled graph states grows exactly as $d=2^n$, because the core stabilizer about $n$ qubits contains $1/4^n$ elements of the original stabilizer group for $N-n$ qubits.

 We will again refer to the structure of IDs.  An ID is a set of observables from within an $N$-qubit stabilizer group whose overall product is $\pm I$ (in the space of all $N$ qubits), and indeed the entire stabilizer group is itself always an ID.  An ID is necessarily composed of {\it single-qubit products} (SQPs), which are ordered sets of single-qubit Pauli observables (and the single-qubit identity) whose overall product is $\pm I$ (for that qubit).  All of the observables in the stabilizer mutually commute, which restricts the relative order of the SQPs (though obviously not the order of the $N$-qubit observables).  We will construct the most general SQP that can belong to a maximally entangled stabilizer, and prove that it has an equal mix of elements $\{Z,X,Y,I\}$.

The proof is straightforward.  To begin, we note that the most fundamental SQP in any entangled stabilizer must contain $\{Z,X\}$ in order for it to be bonded to other qubits, and this in turn will generate $\{Y,I\}$.  We also note that an $N$-qubit stabilizer with $m$ independent elements generates $2^{m}$ $N$-qubit observables, including the $N$-qubit identity.  It therefore follows that for $m=2$, all $N$ SQPs of the stabilizer must contain each of $\{Z,X,Y,I\}$ exactly once.

For $m = 3$ the situation becomes slightly more subtle.  The SQPs each contain 8 elements, and so at least one of $\{Z,X,Y,I\}$ must be repeated - possibly several times.  Let us consider the case where we suppose there must be just one more element $I$ added to the SQP.  First, let $\{A,B,C,D\}$ be arbitrary Pauli observables (in practice these are somewhat constrained, but that is not relevant to this argument), such that we begin with the stabilizer elements above the line in Fig. \ref{SQP23}, and add the next one.  We then generate all of the new product observables which must belong to the complete stabilizer group.
\begin{figure}
\centering
\begin{tabular}{ccc}
$Z$ & $\otimes$ & $A$  \\
$X$ & $\otimes$ & $B$  \\
$Y$ & $\otimes$ & $C$  \\
$I$ & $\otimes$ & $I$  \\
\hline
$I$ & $\otimes$ & $D$  \\
\hline
$Z$ & $\otimes$ & $AD$  \\
$X$ & $\otimes$ & $BD$  \\
$Y$ & $\otimes$ & $CD$  \\
\end{tabular}
\caption{Generation of entangled SQP for $m=3$.}\label{SQP23}
\end{figure}

The only constraint on the above choices is that $D \neq I$, such that we are truly adding a new observable to the set.  It then follows that $AD \neq A$, $BD \neq B$, and $CD \neq C$, which shows that these 8 are each distinct and thus define the complete SQP.  It is straightforward to verify that we obtain the same result if we begin by adding an extra element $Z$,$X$, or $Y$ instead of $I$.  We have therefore shown that for $m=3$, the SQP contains each of $\{Z,X,Y,I\}$ exactly twice.

Reorganizing Fig. \ref{SQP23}, and now adding subscripts to our arbitrary Pauli observables, such that $A_1 \neq A_2$, $B_1 \neq B_2$, $C_1 \neq C_2$, $D_1 \neq D_2$, and $D_i \neq I$, we can repeat the same procedure to obtain the SQPs for $m=4$, as shown in Fig. \ref{SQP34}.
\begin{figure}
\centering
\begin{tabular}{ccc}
$Z$ & $\otimes$ & $A_1$ \\
$Z$ & $\otimes$ & $A_2$  \\
$X$ & $\otimes$ & $B_1$  \\
$X$ & $\otimes$ & $B_2$  \\
$Y$ & $\otimes$ & $C_1$  \\
$Y$ & $\otimes$ & $C_2$  \\
$I$ & $\otimes$ & $D_1$  \\
$I$ & $\otimes$ & $I$  \\
\hline
$I$ & $\otimes$ & $D_2$ \\
\hline
$Z$ & $\otimes$ & $A_1D_2$ \\
$Z$ & $\otimes$ & $A_2D_2$  \\
$X$ & $\otimes$ & $B_1D_2$  \\
$X$ & $\otimes$ & $B_2D_2$  \\
$Y$ & $\otimes$ & $C_1D_2$  \\
$Y$ & $\otimes$ & $C_2D_2$  \\
$I$ & $\otimes$ & $D_1D_2$  \\
\end{tabular}
\caption{Generation of entangled SQP for $m=4$.}\label{SQP34}
\end{figure}
Again, it is straightforward to verify that all 16 elements here are distinct, and thus form the complete SQP.  From here it is easy to see that the number of each element in the SQP doubles every time $m$ is increased by one, and thus the SQP remains an equal mix of $\{Z,X,Y,I\}$ for all $m$, Q.E.D.

This argument holds for any stabilizer group in which every qubit is entangled (bonded) with at least one other qubit, even if the set is not maximally entangled.  Examples of this property of entangled stabilizer SQPs can be seen in nearly all of the figures throughout this paper.

We finally note that an analogous argument shows that stabilizer SQPs that cannot be bonded (unentangled qubits) are always an equal mix of $I$ and {\it one} Pauli observable $Z$, $X$, or $Y$.  Thus the core stabilizer about such a qubit has twice as many elements as does the core stabilizer about an entangled qubit, and defines the same rank-1 projector that would be obtained by measuring that qubit.

\begin{figure}[h!]
\caption{The truncation process for a projective measurement $Z_1$ on the 4-qubit Cluster state, leaving the remaining qubits in a 3-qubit GHZ state.}
\centering
\resizebox{0.8\textwidth}{!}{\begin{minipage}{\textwidth}
\subfloat[][$P_{\pm Z_1} |\psi_{Cluster^4}\rangle$]{
\begin{tabular}{c|cccc}
$\lambda_{1}$ & $\textbf{Z}$ & $Y$ & $X$ & $Y$ \\
$\lambda_{2}$ & $\textbf{Z}$ & $Y$ & $Y$ & $X$ \\
\st{$\lambda_{3}$} & \st{$X$} & \st{$X$} & \st{$I$} & \st{$I$} \\
\st{$\lambda_{4}$} & \st{$X$} & \st{$I$} & \st{$X$} & \st{$X$} \\
$\lambda_{5}$ & $\textbf{I}$ & $X$ & $Y$ & $Y$ \\
$\lambda_{6}$ & $\textbf{I}$ & $I$ & $Z$ & $Z$ \\
\st{$\lambda_{7}$} & \st{$Y$} & \st{$Z$} & \st{$X$} & \st{$Y$} \\
\st{$\lambda_{8}$} & \st{$Y$} & \st{$Y$} & \st{$I$} & \st{$Z$} \\
$\lambda_{9}$ & $\textbf{Z}$ & $Z$ & $Z$ & $I$ \\
\st{$\lambda_{10}$} & \st{$Y$} & \st{$Z$} & \st{$Y$} & \st{$X$} \\
\st{$\lambda_{11}$} & \st{$Y$} & \st{$Y$} & \st{$Z$} & \st{$I$} \\
$\lambda_{12}$ & $\textbf{Z}$ & $Z$ & $I$ & $Z$ \\
$\lambda_{13}$ & $\textbf{I}$ & $X$ & $X$ & $X$ \\
\st{$\lambda_{14}$} & \st{$X$} & \st{$I$} & \st{$Y$} & \st{$Y$} \\
\st{$\lambda_{15}$} & \st{$X$} & \st{$X$} & \st{$Z$} & \st{$Z$} \\
\end{tabular}\label{Cluster_PZ}}
\qquad
\subfloat[][$|\psi_{GHZ^3}\rangle$]{
\begin{tabular}{c|ccc}
$\pm \lambda_{1}$ & $Y$ & $X$ & $Y$ \\
$\pm \lambda_{2}$ & $Y$ & $Y$ & $X$ \\
$\lambda_{5}$ & $X$ & $Y$ & $Y$ \\
$\lambda_{6}$ & $I$ & $Z$ & $Z$ \\
$\pm \lambda_{9}$ & $Z$ & $Z$ & $I$ \\
$\pm \lambda_{12}$ & $Z$ & $I$ & $Z$ \\
$\lambda_{13}$ & $X$ & $X$ & $X$ \\
\end{tabular}\label{Cluster_APZ}}
\end{minipage}}
\label{ClusterPZ}
\end{figure}
\begin{figure}[h!]
\caption{The truncation process for a projective measurement $X_1$ on the 4-qubit Cluster state, leaving the remaining qubits in a partially separable state.}
\centering
\resizebox{0.8\textwidth}{!}{\begin{minipage}{\textwidth}
\subfloat[][$P_{\pm X_1} |\psi_{Cluster^4}\rangle$]{
\begin{tabular}{c|cccc}
\st{$\lambda_{1}$} & \st{$Z$} & \st{$Y$} & \st{$X$} & \st{$Y$} \\
\st{$\lambda_{2}$} & \st{$Z$} & \st{$Y$} & \st{$Y$} & \st{$X$} \\
$\lambda_{3}$ & $\textbf{X}$ & $X$ & $I$ & $I$ \\
$\lambda_{4}$ & $\textbf{X}$ & $I$ & $X$ & $X$ \\
$\lambda_{5}$ & $\textbf{I}$ & $X$ & $Y$ & $Y$ \\
$\lambda_{6}$ & $\textbf{I}$ & $I$ & $Z$ & $Z$ \\
\st{$\lambda_{7}$} & \st{$Y$} & \st{$Z$} & \st{$X$} & \st{$Y$} \\
\st{$\lambda_{8}$} & \st{$Y$} & \st{$Y$} & \st{$I$} & \st{$Z$} \\
\st{$\lambda_{9}$} & \st{$Z$} & \st{$Z$} & \st{$Z$} & \st{$I$} \\
\st{$\lambda_{10}$} & \st{$Y$} & \st{$Z$} & \st{$Y$} & \st{$X$} \\
\st{$\lambda_{11}$} & \st{$Y$} & \st{$Y$} & \st{$Z$} & \st{$I$} \\
\st{$\lambda_{12}$} & \st{$Z$} & \st{$Z$} & \st{$I$} & \st{$Z$} \\
$\lambda_{13}$ & $\textbf{I}$ & $X$ & $X$ & $X$ \\
$\lambda_{14}$ & $\textbf{X}$ & $I$ & $Y$ & $Y$ \\
$\lambda_{15}$ & $\textbf{X}$ & $X$ & $Z$ & $Z$ \\
\end{tabular}\label{Cluster_PX}}
\qquad
\subfloat[][$|X\rangle \otimes |\psi_{Bell}\rangle$]{
\begin{tabular}{c|c:cc}
$\pm \lambda_{3}$ & $X$ & $I$ & $I$ \\
$\pm \lambda_{4}$ & $I$ & $X$ & $X$ \\
$\lambda_{5}$ & $X$ & $Y$ & $Y$ \\
$\lambda_{6}$ & $I$ & $Z$ & $Z$ \\
$\lambda_{13}$ & $X$ & $X$ & $X$ \\
$\pm \lambda_{14}$ & $I$ & $Y$ & $Y$ \\
$\pm \lambda_{15}$ & $X$ & $Z$ & $Z$ \\
\end{tabular}\label{Cluster_APX}}
\end{minipage}}
\label{ClusterPX}
\end{figure}
\begin{figure}[h!]
\caption{The truncation process for a projective measurement $Z_1$ on the 5-qubit Cluster state.}
\centering
\resizebox{0.7\textwidth}{!}{\begin{minipage}{\textwidth}
\subfloat[][$P_{\pm Z_1} |\psi_{Cluster^5}\rangle$]{
\begin{tabular}{c|ccccc}
\st{$\lambda_{1}$} & \st{$Y$} & \st{$Y$} & \st{$Y$} & \st{$Y$} & \st{$Z$} \\
\st{$\lambda_{2}$} & \st{$Y$} & \st{$Y$} & \st{$Z$} & \st{$I$} & \st{$X$} \\
$\lambda_{3}$ & $\textbf{Z}$ & $Z$ & $I$ & $I$ & $I$ \\
$\lambda_{4}$ & $\textbf{Z}$ & $I$ & $Y$ & $Y$ & $X$ \\
$\lambda_{5}$ & $\textbf{I}$ & $Z$ & $I$ & $Z$ & $Z$ \\
$\lambda_{6}$ & $\textbf{I}$ & $I$ & $X$ & $Y$ & $Y$ \\
\st{$\lambda_{7}$} & \st{$X$} & \st{$X$} & \st{$Y$} & \st{$Y$} & \st{$Z$} \\
\st{$\lambda_{8}$} & \st{$X$} & \st{$Y$} & \st{$I$} & \st{$I$} & \st{$Y$} \\
\st{$\lambda_{9}$} & \st{$Y$} & \st{$X$} & \st{$Y$} & \st{$X$} & \st{$I$} \\
\st{$\lambda_{10}$} & \st{$X$} & \st{$X$} & \st{$Z$} & \st{$I$} & \st{$X$} \\
\st{$\lambda_{11}$} & \st{$X$} & \st{$Y$} & \st{$X$} & \st{$Y$} & \st{$I$} \\
\st{$\lambda_{12}$} & \st{$Y$} & \st{$X$} & \st{$Z$} & \st{$Z$} & \st{$Y$} \\
$\lambda_{13}$ & $\textbf{I}$ & $Z$ & $Y$ & $Y$ & $X$ \\
$\lambda_{14}$ & $\textbf{Z}$ & $I$ & $I$ & $Z$ & $Z$ \\
$\lambda_{15}$ & $\textbf{Z}$ & $Z$ & $Y$ & $X$ & $Y$ \\
$\lambda_{16}$ & $\textbf{Z}$ & $Z$ & $X$ & $Y$ & $Y$ \\
$\lambda_{17}$ & $\textbf{Z}$ & $I$ & $Z$ & $I$ & $Z$ \\
$\lambda_{18}$ & $\textbf{I}$ & $Z$ & $X$ & $X$ & $X$ \\
\st{$\lambda_{19}$} & \st{$Y$} & \st{$X$} & \st{$I$} & \st{$I$} & \st{$Y$} \\
\st{$\lambda_{20}$} & \st{$X$} & \st{$Y$} & \st{$Y$} & \st{$X$} & \st{$I$} \\
\st{$\lambda_{21}$} & \st{$X$} & \st{$X$} & \st{$I$} & \st{$Z$} & \st{$X$} \\
\st{$\lambda_{22}$} & \st{$Y$} & \st{$X$} & \st{$X$} & \st{$Y$} & \st{$I$} \\
\st{$\lambda_{23}$} & \st{$X$} & \st{$Y$} & \st{$Z$} & \st{$Z$} & \st{$Y$} \\
\st{$\lambda_{24}$} & \st{$X$} & \st{$X$} & \st{$X$} & \st{$X$} & \st{$Z$} \\
$\lambda_{25}$ & $\textbf{I}$ & $I$ & $Y$ & $X$ & $Y$ \\
$\lambda_{26}$ & $\textbf{I}$ & $Z$ & $Z$ & $I$ & $Z$ \\
$\lambda_{27}$ & $\textbf{Z}$ & $I$ & $X$ & $X$ & $X$ \\
$\lambda_{28}$ & $\textbf{Z}$ & $Z$ & $Z$ & $Z$ & $I$ \\
\st{$\lambda_{29}$} & \st{$Y$} & \st{$Y$} & \st{$I$} & \st{$Z$} & \st{$X$} \\
\st{$\lambda_{30}$} & \st{$Y$} & \st{$Y$} & \st{$X$} & \st{$X$} & \st{$Z$} \\
$\lambda_{31}$ & $\textbf{I}$ & $I$ & $Z$ & $Z$ & $I$ \\
\end{tabular}\label{Cluster5_PZ1}}
\qquad
\subfloat[][$|Z\rangle \otimes |\psi_{GHZ^3}\rangle$]{
\begin{tabular}{c|c:ccc}
$\pm \lambda_{3}$ & $Z$ & $I$ & $I$ & $I$ \\
$\pm \lambda_{4}$ & $I$ & $Y$ & $Y$ & $X$ \\
$\lambda_{5}$ & $Z$ & $I$ & $Z$ & $Z$ \\
$\lambda_{6}$ & $I$ & $X$ & $Y$ & $Y$ \\
$\lambda_{13}$ & $Z$ & $Y$ & $Y$ & $X$ \\
$\pm \lambda_{14}$ & $I$ & $I$ & $Z$ & $Z$ \\
$\pm \lambda_{15}$ & $Z$ & $Y$ & $X$ & $Y$ \\
$\pm \lambda_{16}$ & $Z$ & $X$ & $Y$ & $Y$ \\
$\pm \lambda_{17}$ & $I$ & $Z$ & $I$ & $Z$ \\
$\lambda_{18}$ & $Z$ & $X$ & $X$ & $X$ \\
$\lambda_{25}$ & $I$ & $Y$ & $X$ & $Y$ \\
$\lambda_{26}$ & $Z$ & $Z$ & $I$ & $Z$ \\
$\pm \lambda_{27}$ & $I$ & $X$ & $X$ & $X$ \\
$\pm \lambda_{28}$ & $Z$ & $Z$ & $Z$ & $I$ \\
$\lambda_{31}$ & $I$ & $Z$ & $Z$ & $I$ \\
\end{tabular}\label{Cluster5_APZ1}}
\end{minipage}}
\label{Cluster5PZ1}
\end{figure}
\begin{figure}[h!]
\caption{The truncation process for a projective measurement $X_1$ on the 5-qubit Cluster state.}
\centering
\resizebox{0.7\textwidth}{!}{\begin{minipage}{\textwidth}
\subfloat[][$P_{\pm X_1} |\psi_{Cluster^5}\rangle$]{
\begin{tabular}{c|ccccc}
\st{$\lambda_{1}$} & \st{$Y$} & \st{$Y$} & \st{$Y$} & \st{$Y$} & \st{$Z$} \\
\st{$\lambda_{2}$} & \st{$Y$} & \st{$Y$} & \st{$Z$} & \st{$I$} & \st{$X$} \\
\st{$\lambda_{3}$} & \st{$Z$} & \st{$Z$} & \st{$I$} & \st{$I$} & \st{$I$} \\
\st{$\lambda_{4}$} & \st{$Z$} & \st{$I$} & \st{$Y$} & \st{$Y$} & \st{$X$} \\
$\lambda_{5}$ & $\textbf{I}$ & $Z$ & $I$ & $Z$ & $Z$ \\
$\lambda_{6}$ & $\textbf{I}$ & $I$ & $X$ & $Y$ & $Y$ \\
$\lambda_{7}$ & $\textbf{X}$ & $X$ & $Y$ & $Y$ & $Z$ \\
$\lambda_{8}$ & $\textbf{X}$ & $Y$ & $I$ & $I$ & $Y$ \\
\st{$\lambda_{9}$} & \st{$Y$} & \st{$X$} & \st{$Y$} & \st{$X$} & \st{$I$} \\
$\lambda_{10}$ & $\textbf{X}$ & $X$ & $Z$ & $I$ & $X$ \\
$\lambda_{11}$ & $\textbf{X}$ & $Y$ & $X$ & $Y$ & $I$ \\
\st{$\lambda_{12}$} & \st{$Y$} & \st{$X$} & \st{$Z$} & \st{$Z$} & \st{$Y$} \\
$\lambda_{13}$ & $\textbf{I}$ & $Z$ & $Y$ & $Y$ & $X$ \\
\st{$\lambda_{14}$} & \st{$Z$} & \st{$I$} & \st{$I$} & \st{$Z$} & \st{$Z$} \\
\st{$\lambda_{15}$} & \st{$Z$} & \st{$Z$} & \st{$Y$} & \st{$X$} & \st{$Y$} \\
\st{$\lambda_{16}$} & \st{$Z$} & \st{$Z$} & \st{$X$} & \st{$Y$} & \st{$Y$} \\
\st{$\lambda_{17}$} & \st{$Z$} & \st{$I$} & \st{$Z$} & \st{$I$} & \st{$Z$} \\
$\lambda_{18}$ & $\textbf{I}$ & $Z$ & $X$ & $X$ & $X$ \\
\st{$\lambda_{19}$} & \st{$Y$} & \st{$X$} & \st{$I$} & \st{$I$} & \st{$Y$} \\
$\lambda_{20}$ & $\textbf{X}$ & $Y$ & $Y$ & $X$ & $I$ \\
$\lambda_{21}$ & $\textbf{X}$ & $X$ & $I$ & $Z$ & $X$ \\
\st{$\lambda_{22}$} & \st{$Y$} & \st{$X$} & \st{$X$} & \st{$Y$} & \st{$I$} \\
$\lambda_{23}$ & $\textbf{X}$ & $Y$ & $Z$ & $Z$ & $Y$ \\
$\lambda_{24}$ & $\textbf{X}$ & $X$ & $X$ & $X$ & $Z$ \\
$\lambda_{25}$ & $\textbf{I}$ & $I$ & $Y$ & $X$ & $Y$ \\
$\lambda_{26}$ & $\textbf{I}$ & $Z$ & $Z$ & $I$ & $Z$ \\
\st{$\lambda_{27}$} & \st{$Z$} & \st{$I$} & \st{$X$} & \st{$X$} & \st{$X$} \\
\st{$\lambda_{28}$} & \st{$Z$} & \st{$Z$} & \st{$Z$} & \st{$Z$} & \st{$I$} \\
\st{$\lambda_{29}$} & \st{$Y$} & \st{$Y$} & \st{$I$} & \st{$Z$} & \st{$X$} \\
\st{$\lambda_{30}$} & \st{$Y$} & \st{$Y$} & \st{$X$} & \st{$X$} & \st{$Z$} \\
$\lambda_{31}$ & $\textbf{I}$ & $I$ & $Z$ & $Z$ & $I$ \\
\end{tabular}\label{Cluster5_PX1}}
\qquad
\subfloat[][$|\psi_{Cluster^4}\rangle$]{
\begin{tabular}{c|cccc}
$\lambda_{5}$ & $Z$ & $I$ & $Z$ & $Z$ \\
$\lambda_{6}$ & $I$ & $X$ & $Y$ & $Y$ \\
$\pm \lambda_{7}$ & $X$ & $Y$ & $Y$ & $Z$ \\
$\pm \lambda_{8}$ & $Y$ & $I$ & $I$ & $Y$ \\
$\pm \lambda_{10}$ & $X$ & $Z$ & $I$ & $X$ \\
$\pm \lambda_{11}$ & $Y$ & $X$ & $Y$ & $I$ \\
$\lambda_{13}$ & $Z$ & $Y$ & $Y$ & $X$ \\
$\lambda_{18}$ & $Z$ & $X$ & $X$ & $X$ \\
$\pm \lambda_{20}$ & $Y$ & $Y$ & $X$ & $I$ \\
$\pm \lambda_{21}$ & $X$ & $I$ & $Z$ & $X$ \\
$\pm \lambda_{23}$ & $Y$ & $Z$ & $Z$ & $Y$ \\
$\pm \lambda_{24}$ & $X$ & $X$ & $X$ & $Z$ \\
$\lambda_{25}$ & $I$ & $Y$ & $X$ & $Y$ \\
$\lambda_{26}$ & $Z$ & $Z$ & $I$ & $Z$ \\
$\lambda_{31}$ & $I$ & $Z$ & $Z$ & $I$ \\
\end{tabular}\label{Cluster5_APX1}}
\end{minipage}}
\label{Cluster5PX1}
\end{figure}

\begin{figure}[h!]
\caption{The truncation process to obtain the core state about qubit 1 of the 5-qubit cluster state.  Note that the core space is not maximally entangled, but contains states with either 3-party or 4-party entanglement.}
\centering
\resizebox{0.7\textwidth}{!}{\begin{minipage}{\textwidth}
\subfloat[][core$_1|\psi_{Cluster^5}\rangle$]{
\begin{tabular}{c|ccccc}
\st{$\lambda_{1}$} & \st{$Y$} & \st{$Y$} & \st{$Y$} & \st{$Y$} & \st{$Z$} \\
\st{$\lambda_{2}$} & \st{$Y$} & \st{$Y$} & \st{$Z$} & \st{$I$} & \st{$X$} \\
\st{$\lambda_{3}$} & \st{$Z$} & \st{$Z$} & \st{$I$} & \st{$I$} & \st{$I$} \\
\st{$\lambda_{4}$} & \st{$Z$} & \st{$I$} & \st{$Y$} & \st{$Y$} & \st{$X$} \\
$\lambda_{5}$ & $\textbf{I}$ & $Z$ & $I$ & $Z$ & $Z$ \\
$\lambda_{6}$ & $\textbf{I}$ & $I$ & $X$ & $Y$ & $Y$ \\
\st{$\lambda_{7}$} & \st{$X$} & \st{$X$} & \st{$Y$} & \st{$Y$} & \st{$Z$} \\
\st{$\lambda_{8}$} & \st{$X$} & \st{$Y$} & \st{$I$} & \st{$I$} & \st{$Y$} \\
\st{$\lambda_{9}$} & \st{$Y$} & \st{$X$} & \st{$Y$} & \st{$X$} & \st{$I$} \\
\st{$\lambda_{10}$} & \st{$X$} & \st{$X$} & \st{$Z$} & \st{$I$} & \st{$X$} \\
\st{$\lambda_{11}$} & \st{$X$} & \st{$Y$} & \st{$X$} & \st{$Y$} & \st{$I$} \\
\st{$\lambda_{12}$} & \st{$Y$} & \st{$X$} & \st{$Z$} & \st{$Z$} & \st{$Y$} \\
$\lambda_{13}$ & $\textbf{I}$ & $Z$ & $Y$ & $Y$ & $X$ \\
\st{$\lambda_{14}$} & \st{$Z$} & \st{$I$} & \st{$I$} & \st{$Z$} & \st{$Z$} \\
\st{$\lambda_{15}$} & \st{$Z$} & \st{$Z$} & \st{$Y$} & \st{$X$} & \st{$Y$} \\
\st{$\lambda_{16}$} & \st{$Z$} & \st{$Z$} & \st{$X$} & \st{$Y$} & \st{$Y$} \\
\st{$\lambda_{17}$} & \st{$Z$} & \st{$I$} & \st{$Z$} & \st{$I$} & \st{$Z$} \\
$\lambda_{18}$ & $\textbf{I}$ & $Z$ & $X$ & $X$ & $X$ \\
\st{$\lambda_{19}$} & \st{$Y$} & \st{$X$} & \st{$I$} & \st{$I$} & \st{$Y$} \\
\st{$\lambda_{20}$} & \st{$X$} & \st{$Y$} & \st{$Y$} & \st{$X$} & \st{$I$} \\
\st{$\lambda_{21}$} & \st{$X$} & \st{$X$} & \st{$I$} & \st{$Z$} & \st{$X$} \\
\st{$\lambda_{22}$} & \st{$Y$} & \st{$X$} & \st{$X$} & \st{$Y$} & \st{$I$} \\
\st{$\lambda_{23}$} & \st{$X$} & \st{$Y$} & \st{$Z$} & \st{$Z$} & \st{$Y$} \\
\st{$\lambda_{24}$} & \st{$X$} & \st{$X$} & \st{$X$} & \st{$X$} & \st{$Z$} \\
$\lambda_{25}$ & $\textbf{I}$ & $I$ & $Y$ & $X$ & $Y$ \\
$\lambda_{26}$ & $\textbf{I}$ & $Z$ & $Z$ & $I$ & $Z$ \\
\st{$\lambda_{27}$} & \st{$Z$} & \st{$I$} & \st{$X$} & \st{$X$} & \st{$X$} \\
\st{$\lambda_{28}$} & \st{$Z$} & \st{$Z$} & \st{$Z$} & \st{$Z$} & \st{$I$} \\
\st{$\lambda_{29}$} & \st{$Y$} & \st{$Y$} & \st{$I$} & \st{$Z$} & \st{$X$} \\
\st{$\lambda_{30}$} & \st{$Y$} & \st{$Y$} & \st{$X$} & \st{$X$} & \st{$Z$} \\
$\lambda_{31}$ & $\textbf{I}$ & $I$ & $Z$ & $Z$ & $I$ \\
\end{tabular}\label{Cluster5State}}
\qquad
\subfloat[][This rank-2 core space contains, and is spanned by, the states of Figs. \ref{Cluster5_APZ1} and \ref{Cluster5_APX1}.]{
\begin{tabular}{c|c:ccc}
$\lambda_{5}$ & $Z$ & $I$ & $Z$ & $Z$ \\
$\lambda_{6}$ & $I$ & $X$ & $Y$ & $Y$ \\
$\lambda_{13}$ & $Z$ & $Y$ & $Y$ & $X$ \\
$\lambda_{18}$ & $Z$ & $X$ & $X$ & $X$ \\
$\lambda_{25}$ & $I$ & $Y$ & $X$ & $Y$ \\
$\lambda_{26}$ & $Z$ & $Z$ & $I$ & $Z$ \\
$\lambda_{31}$ & $I$ & $Z$ & $Z$ & $I$ \\
\end{tabular}\label{Cluster5Core}}
\end{minipage}}
\label{ClusterCore}
\end{figure}

\begin{figure}[h!]
\caption{The truncation process for a projective measurement $Z_5$ on the 5-qubit Cluster state.}
\centering
\resizebox{0.7\textwidth}{!}{\begin{minipage}{\textwidth}
\subfloat[][$P_{\pm Z_5} |\psi_{Cluster^5}\rangle$]{
\begin{tabular}{c|ccccc}
$\lambda_{1}$ & $Y$ & $Y$ & $Y$ & $Y$ & $\textbf{Z}$ \\
\st{$\lambda_{2}$} & \st{$Y$} & \st{$Y$} & \st{$Z$} & \st{$I$} & \st{$X$} \\
$\lambda_{3}$ & $Z$ & $Z$ & $I$ & $I$ & $\textbf{I}$ \\
\st{$\lambda_{4}$} & \st{$Z$} & \st{$I$} & \st{$Y$} & \st{$Y$} & \st{$X$} \\
$\lambda_{5}$ & $I$ & $Z$ & $I$ & $Z$ & $\textbf{Z}$ \\
\st{$\lambda_{6}$} & \st{$I$} & \st{$I$} & \st{$X$} & \st{$Y$} & \st{$Y$} \\
$\lambda_{7}$ & $X$ & $X$ & $Y$ & $Y$ & $\textbf{Z}$ \\
\st{$\lambda_{8}$} & \st{$X$} & \st{$Y$} & \st{$I$} & \st{$I$} & \st{$Y$} \\
$\lambda_{9}$ & $Y$ & $X$ & $Y$ & $X$ & $\textbf{I}$ \\
\st{$\lambda_{10}$} & \st{$X$} & \st{$X$} & \st{$Z$} & \st{$I$} & \st{$X$} \\
$\lambda_{11}$ & $X$ & $Y$ & $X$ & $Y$ & $\textbf{I}$ \\
\st{$\lambda_{12}$} & \st{$Y$} & \st{$X$} & \st{$Z$} & \st{$Z$} & \st{$Y$} \\
\st{$\lambda_{13}$} & \st{$I$} & \st{$Z$} & \st{$Y$} & \st{$Y$} & \st{$X$} \\
$\lambda_{14}$ & $Z$ & $I$ & $I$ & $Z$ & $\textbf{Z}$ \\
\st{$\lambda_{15}$} & \st{$Z$} & \st{$Z$} & \st{$Y$} & \st{$X$} & \st{$Y$} \\
\st{$\lambda_{16}$} & \st{$Z$} & \st{$Z$} & \st{$X$} & \st{$Y$} & \st{$Y$} \\
$\lambda_{17}$ & $Z$ & $I$ & $Z$ & $I$ & $\textbf{Z}$ \\
\st{$\lambda_{18}$} & \st{$I$} & \st{$Z$} & \st{$X$} & \st{$X$} & \st{$X$} \\
\st{$\lambda_{19}$} & \st{$Y$} & \st{$X$} & \st{$I$} & \st{$I$} & \st{$Y$} \\
$\lambda_{20}$ & $X$ & $Y$ & $Y$ & $X$ & $\textbf{I}$ \\
\st{$\lambda_{21}$} & \st{$X$} & \st{$X$} & \st{$I$} & \st{$Z$} & \st{$X$} \\
$\lambda_{22}$ & $Y$ & $X$ & $X$ & $Y$ & $\textbf{I}$ \\
\st{$\lambda_{23}$} & \st{$X$} & \st{$Y$} & \st{$Z$} & \st{$Z$} & \st{$Y$} \\
$\lambda_{24}$ & $X$ & $X$ & $X$ & $X$ & $\textbf{Z}$ \\
\st{$\lambda_{25}$} & \st{$I$} & \st{$I$} & \st{$Y$} & \st{$X$} & \st{$Y$} \\
$\lambda_{26}$ & $I$ & $Z$ & $Z$ & $I$ & $\textbf{Z}$ \\
\st{$\lambda_{27}$} & \st{$Z$} & \st{$I$} & \st{$X$} & \st{$X$} & \st{$X$} \\
$\lambda_{28}$ & $Z$ & $Z$ & $Z$ & $Z$ & $\textbf{I}$ \\
\st{$\lambda_{29}$} & \st{$Y$} & \st{$Y$} & \st{$I$} & \st{$Z$} & \st{$X$} \\
$\lambda_{30}$ & $Y$ & $Y$ & $X$ & $X$ & $\textbf{Z}$ \\
$\lambda_{31}$ & $I$ & $I$ & $Z$ & $Z$ & $\textbf{I}$ \\
\end{tabular}\label{Cluster5_PZ5}}
\qquad
\subfloat[][$|\psi_{GHZ^4}\rangle$]{
\begin{tabular}{c|cccc}
$\pm \lambda_{1}$ & $Y$ & $Y$ & $Y$ & $Y$ \\
$\lambda_{3}$ & $Z$ & $Z$ & $I$ & $I$ \\
$\pm \lambda_{5}$ & $I$ & $Z$ & $I$ & $Z$ \\
$\pm \lambda_{7}$ & $X$ & $X$ & $Y$ & $Y$ \\
$\lambda_{9}$ & $Y$ & $X$ & $Y$ & $X$ \\
$\lambda_{11}$ & $X$ & $Y$ & $X$ & $Y$ \\
$\pm \lambda_{14}$ & $Z$ & $I$ & $I$ & $Z$ \\
$\pm \lambda_{17}$ & $Z$ & $I$ & $Z$ & $I$ \\
$\lambda_{20}$ & $X$ & $Y$ & $Y$ & $X$ \\
$\lambda_{22}$ & $Y$ & $X$ & $X$ & $Y$ \\
$\pm \lambda_{24}$ & $X$ & $X$ & $X$ & $X$ \\
$\pm \lambda_{26}$ & $I$ & $Z$ & $Z$ & $I$ \\
$\lambda_{28}$ & $Z$ & $Z$ & $Z$ & $Z$ \\
$\pm \lambda_{30}$ & $Y$ & $Y$ & $X$ & $X$ \\
$\lambda_{31}$ & $I$ & $I$ & $Z$ & $Z$ \\
\end{tabular}\label{Cluster5_APZ5}}
\end{minipage}}
\label{Cluster5PZ5}
\end{figure}
\begin{figure}[h!]
\caption{The truncation process for a projective measurement $Y_5$ on the 5-qubit Cluster state.}
\centering
\resizebox{0.7\textwidth}{!}{\begin{minipage}{\textwidth}
\subfloat[][$P_{\pm Y_5} |\psi_{Cluster^5}\rangle$]{
\begin{tabular}{c|ccccc}
\st{$\lambda_{1}$} & \st{$Y$} & \st{$Y$} & \st{$Y$} & \st{$Y$} & \st{$Z$} \\
\st{$\lambda_{2}$} & \st{$Y$} & \st{$Y$} & \st{$Z$} & \st{$I$} & \st{$X$} \\
$\lambda_{3}$ & $Z$ & $Z$ & $I$ & $I$ & $\textbf{I}$ \\
\st{$\lambda_{4}$} & \st{$Z$} & \st{$I$} & \st{$Y$} & \st{$Y$} & \st{$X$} \\
\st{$\lambda_{5}$} & \st{$I$} & \st{$Z$} & \st{$I$} & \st{$Z$} & \st{$Z$} \\
$\lambda_{6}$ & $I$ & $I$ & $X$ & $Y$ & $\textbf{Y}$ \\
\st{$\lambda_{7}$} & \st{$X$} & \st{$X$} & \st{$Y$} & \st{$Y$} & \st{$Z$} \\
$\lambda_{8}$ & $X$ & $Y$ & $I$ & $I$ & $\textbf{Y}$ \\
$\lambda_{9}$ & $Y$ & $X$ & $Y$ & $X$ & $\textbf{I}$ \\
\st{$\lambda_{10}$} & \st{$X$} & \st{$X$} & \st{$Z$} & \st{$I$} & \st{$X$} \\
$\lambda_{11}$ & $X$ & $Y$ & $X$ & $Y$ & $\textbf{I}$ \\
$\lambda_{12}$ & $Y$ & $X$ & $Z$ & $Z$ & $\textbf{Y}$ \\
\st{$\lambda_{13}$} & \st{$I$} & \st{$Z$} & \st{$Y$} & \st{$Y$} & \st{$X$} \\
\st{$\lambda_{14}$} & \st{$Z$} & \st{$I$} & \st{$I$} & \st{$Z$} & \st{$Z$} \\
$\lambda_{15}$ & $Z$ & $Z$ & $Y$ & $X$ & $\textbf{Y}$ \\
$\lambda_{16}$ & $Z$ & $Z$ & $X$ & $Y$ & $\textbf{Y}$ \\
\st{$\lambda_{17}$} & \st{$Z$} & \st{$I$} & \st{$Z$} & \st{$I$} & \st{$Z$} \\
\st{$\lambda_{18}$} & \st{$I$} & \st{$Z$} & \st{$X$} & \st{$X$} & \st{$X$} \\
$\lambda_{19}$ & $Y$ & $X$ & $I$ & $I$ & $\textbf{Y}$ \\
$\lambda_{20}$ & $X$ & $Y$ & $Y$ & $X$ & $\textbf{I}$ \\
\st{$\lambda_{21}$} & \st{$X$} & \st{$X$} & \st{$I$} & \st{$Z$} & \st{$X$} \\
$\lambda_{22}$ & $Y$ & $X$ & $X$ & $Y$ & $\textbf{I}$ \\
$\lambda_{23}$ & $X$ & $Y$ & $Z$ & $Z$ & $\textbf{Y}$ \\
\st{$\lambda_{24}$} & \st{$X$} & \st{$X$} & \st{$X$} & \st{$X$} & \st{$Z$} \\
$\lambda_{25}$ & $I$ & $I$ & $Y$ & $X$ & $\textbf{Y}$ \\
\st{$\lambda_{26}$} & \st{$I$} & \st{$Z$} & \st{$Z$} & \st{$I$} & \st{$Z$} \\
\st{$\lambda_{27}$} & \st{$Z$} & \st{$I$} & \st{$X$} & \st{$X$} & \st{$X$} \\
$\lambda_{28}$ & $Z$ & $Z$ & $Z$ & $Z$ & $\textbf{I}$ \\
\st{$\lambda_{29}$} & \st{$Y$} & \st{$Y$} & \st{$I$} & \st{$Z$} & \st{$X$} \\
\st{$\lambda_{30}$} & \st{$Y$} & \st{$Y$} & \st{$X$} & \st{$X$} & \st{$Z$} \\
$\lambda_{31}$ & $I$ & $I$ & $Z$ & $Z$ & $\textbf{I}$ \\
\end{tabular}\label{Cluster5_PY5}}
\qquad
\subfloat[][$|\psi_{Bell}\rangle \otimes |\psi_{Bell}\rangle$]{
\begin{tabular}{c|cc:cc}
$\lambda_{3}$ & $Z$ & $Z$ & $I$ & $I$ \\
$\pm \lambda_{6}$ & $I$ & $I$ & $X$ & $Y$ \\
$\pm \lambda_{8}$ & $X$ & $Y$ & $I$ & $I$ \\
$\lambda_{9}$ & $Y$ & $X$ & $Y$ & $X$ \\
$\lambda_{11}$ & $X$ & $Y$ & $X$ & $Y$ \\
$\pm \lambda_{12}$ & $Y$ & $X$ & $Z$ & $Z$ \\
$\pm \lambda_{15}$ & $Z$ & $Z$ & $Y$ & $X$ \\
$\pm \lambda_{16}$ & $Z$ & $Z$ & $X$ & $Y$ \\
$\pm \lambda_{19}$ & $Y$ & $X$ & $I$ & $I$ \\
$\lambda_{20}$ & $X$ & $Y$ & $Y$ & $X$ \\
$\lambda_{22}$ & $Y$ & $X$ & $X$ & $Y$ \\
$\pm \lambda_{23}$ & $X$ & $Y$ & $Z$ & $Z$ \\
$\pm \lambda_{25}$ & $I$ & $I$ & $Y$ & $X$ \\
$\lambda_{28}$ & $Z$ & $Z$ & $Z$ & $Z$ \\
$\lambda_{31}$ & $I$ & $I$ & $Z$ & $Z$ \\
\end{tabular}\label{Cluster5_APY5}}
\end{minipage}}
\label{Cluster5PY5}
\end{figure}

\begin{figure}[h!]
\caption{The truncation process to obtain the core state about qubit 1 of the 5-qubit pentagon state.}
\centering
\resizebox{0.7\textwidth}{!}{\begin{minipage}{\textwidth}
\subfloat[][core$_1(|\psi_{Pentagon^5}\rangle$)]{
\begin{tabular}{c|ccccc}
\st{$\lambda_{1}$} & \st{$X$} & \st{$Z$} & \st{$I$} & \st{$I$} & \st{$Z$} \\
\st{$\lambda_{2}$} & \st{$Z$} & \st{$X$} & \st{$Z$} & \st{$I$} & \st{$I$} \\
$\lambda_{3}$ & $\textbf{I}$ & $Z$ & $X$ & $Z$ & $I$ \\
$\lambda_{4}$ & $\textbf{I}$ & $I$ & $Z$ & $X$ & $Z$ \\
\st{$\lambda_{5}$} & \st{$Z$} & \st{$I$} & \st{$I$} & \st{$Z$} & \st{$X$} \\
\st{$\lambda_{6}$} & \st{$Y$} & \st{$Y$} & \st{$Z$} & \st{$I$} & \st{$Z$} \\
\st{$\lambda_{7}$} & \st{$X$} & \st{$I$} & \st{$X$} & \st{$Z$} & \st{$Z$} \\
\st{$\lambda_{8}$} & \st{$X$} & \st{$Z$} & \st{$Z$} & \st{$X$} & \st{$I$} \\
\st{$\lambda_{9}$} & \st{$Y$} & \st{$Z$} & \st{$I$} & \st{$Z$} & \st{$Y$} \\
\st{$\lambda_{10}$} & \st{$Z$} & \st{$Y$} & \st{$Y$} & \st{$Z$} & \st{$I$} \\
\st{$\lambda_{11}$} & \st{$Z$} & \st{$X$} & \st{$I$} & \st{$X$} & \st{$Z$} \\
$\lambda_{12}$ & $\textbf{I}$ & $X$ & $Z$ & $Z$ & $X$ \\
$\lambda_{13}$ & $\textbf{I}$ & $Z$ & $Y$ & $Y$ & $Z$ \\
\st{$\lambda_{14}$} & \st{$Z$} & \st{$Z$} & \st{$X$} & \st{$I$} & \st{$X$} \\
\st{$\lambda_{15}$} & \st{$Z$} & \st{$I$} & \st{$Z$} & \st{$Y$} & \st{$Y$} \\
\st{$\lambda_{16}$} & \st{$Y$} & \st{$X$} & \st{$Y$} & \st{$Z$} & \st{$Z$} \\
\st{$\lambda_{17}$} & \st{$Y$} & \st{$Y$} & \st{$I$} & \st{$X$} & \st{$I$} \\
\st{$\lambda_{18}$} & \st{$X$} & \st{$Y$} & \st{$Z$} & \st{$Z$} & \st{$Y$} \\
\st{$\lambda_{19}$} & \st{$X$} & \st{$I$} & \st{$Y$} & \st{$Y$} & \st{$I$} \\
\st{$\lambda_{20}$} & \st{$Y$} & \st{$I$} & \st{$X$} & \st{$I$} & \st{$Y$} \\
\st{$\lambda_{21}$} & \st{$Y$} & \st{$Z$} & \st{$Z$} & \st{$Y$} & \st{$X$} \\
\st{$\lambda_{22}$} & \st{$Z$} & \st{$Y$} & \st{$X$} & \st{$Y$} & \st{$Z$} \\
$\lambda_{23}$ & $\textbf{I}$ & $Y$ & $Y$ & $I$ & $X$ \\
$\lambda_{24}$ & $\textbf{I}$ & $X$ & $I$ & $Y$ & $Y$ \\
\st{$\lambda_{25}$} & \st{$Z$} & \st{$Z$} & \st{$Y$} & \st{$X$} & \st{$Y$} \\
\st{$\lambda_{26}$} & \st{$Y$} & \st{$X$} & \st{$X$} & \st{$Y$} & \st{$I$} \\
\st{$\lambda_{27}$} & \st{$X$} & \st{$X$} & \st{$Y$} & \st{$I$} & \st{$Y$} \\
\st{$\lambda_{28}$} & \st{$X$} & \st{$Y$} & \st{$I$} & \st{$Y$} & \st{$X$} \\
\st{$\lambda_{29}$} & \st{$Y$} & \st{$I$} & \st{$Y$} & \st{$X$} & \st{$X$} \\
$\lambda_{30}$ & $\textbf{I}$ & $Y$ & $X$ & $X$ & $Y$ \\
\st{$\lambda_{31}$} & \st{$X$} & \st{$X$} & \st{$X$} & \st{$X$} & \st{$X$} \\
\end{tabular}\label{PentagonState}}
\qquad
\subfloat[][Maximally entangled rank-2 projector (cluster space)]{
\begin{tabular}{c|cccc}
$\lambda_{3}$ & $Z$ & $X$ & $Z$ & $I$ \\
$\lambda_{4}$ & $I$ & $Z$ & $X$ & $Z$ \\
$\lambda_{12}$ & $X$ & $Z$ & $Z$ & $X$ \\
$\lambda_{13}$ & $Z$ & $Y$ & $Y$ & $Z$ \\
$\lambda_{23}$ & $Y$ & $Y$ & $I$ & $X$ \\
$\lambda_{24}$ & $X$ & $I$ & $Y$ & $Y$ \\
$\lambda_{30}$ & $Y$ & $X$ & $X$ & $Y$ \\
\end{tabular}\label{cluster4space}}
\end{minipage}}
\label{PentagonCore}
\end{figure}

\end{document}